\documentclass[pre,superscriptaddress,twocolumn,aps,10pt]{revtex4-1}
\pdfoutput=1
\usepackage[colorlinks=true,urlcolor=blue]{hyperref}
\usepackage{amsfonts}
\usepackage{graphicx}
\usepackage{mathrsfs}
\usepackage{amsmath}
\usepackage{xcolor}
\usepackage{bm}
\usepackage{color}
\usepackage{upgreek}

\usepackage{color}
\definecolor{red}{rgb}{0.75,0,0}
\definecolor{blue}{rgb}{0,0,0.75}
\definecolor{green}{rgb}{0,0.5,0}

\newcommand{\dd}{\mathrm{d}}

\DeclareMathOperator{\tr}{tr}

\begin{document}
\title{Mechanical interplay between cell shape and actin cytoskeleton organization}

\author{Koen Schakenraad}
\affiliation{Instituut-Lorentz, Leiden University, P.O. Box 9506, 2300 RA Leiden, The Netherlands}
\affiliation{Mathematical Institute, Leiden University, P.O. Box 9512, 2300 RA Leiden, The Netherlands}
\author{Jeremy Ernst}
\affiliation{Instituut-Lorentz, Leiden University, P.O. Box 9506, 2300 RA Leiden, The Netherlands}
\author{Wim Pomp}
\affiliation{Kamerlingh Onnes-Huygens Laboratory, Leiden University, P.O. Box 9504, 2300 RA Leiden, The Netherlands}
\affiliation{Division of Gene Regulation, The Netherlands Cancer Institute, P.O. Box 90203, 1006 BE Amsterdam, The Netherlands}
\author{Erik H. J. Danen}
\affiliation{Leiden Academic Center for Drug Research, Leiden University, P.O. Box 9502, 2300 RA Leiden, The Netherlands}
\author{Roeland M. H. Merks}
\affiliation{Mathematical Institute, Leiden University, P.O. Box 9512, 2300 RA Leiden, The Netherlands}
\affiliation{Institute of Biology, Leiden University, P.O. Box 9505, 2300 RA Leiden, The Netherlands}
\author{Thomas Schmidt}
\affiliation{Kamerlingh Onnes-Huygens Laboratory, Leiden University, P.O. Box 9504, 2300 RA Leiden, The Netherlands}
\author{Luca Giomi}
\thanks{Corresponding author: giomi@lorentz.leidenuniv.nl}
\affiliation{Instituut-Lorentz, Leiden University, P.O. Box 9506, 2300 RA Leiden, The Netherlands}

\begin{abstract}
We investigate the mechanical interplay between the spatial organization of the actin cytoskeleton and the shape of animal cells adhering on micropillar arrays. Using a combination of analytical work, computer simulations and {\em in vitro} experiments, we demonstrate that the orientation of the stress fibers strongly influences the geometry of the cell edge. In the presence of a uniformly aligned cytoskeleton, the cell edge can be well approximated by elliptical arcs, whose eccentricity reflects the degree of anisotropy of the cell's internal stresses. Upon modeling the actin cytoskeleton as a nematic liquid crystal, we further show that the geometry of the cell edge feeds back on the organization of the stress fibers by altering the length scale at which these are confined. This feedback mechanism is controlled by a dimensionless number, the anchoring number, representing the relative weight of surface-anchoring and bulk-aligning torques. Our model allows to predict both cellular shape and the internal structure of the actin cytoskeleton and is in good quantitative agreement with experiments on fibroblastoid (GD$\upbeta$1,GD$\upbeta$3) and epithelioid (GE$\upbeta$1, GE$\upbeta$3) cells. 
\end{abstract}

\maketitle
\section{Introduction}
Mechanical cues play a vital role in many cellular processes, such as durotaxis \cite{Lo2000,Sochol2011}, cell-cell communication \cite{King2008}, stress-regulated protein expression \cite{Sawada2006} or rigidity-dependent stem cell differentiation \cite{Engler2006,Trappmann2012}. Whereas mechanical forces can directly activate biochemical signaling pathways, via the mechanotransduction machinery \cite{Panciera2017}, their effect is often mediated by the cortical cytoskeleton, which, in turn, affects and can be affected by the geometry of the cell envelope.

By adjusting their shape, cells can sense the mechanical properties of their microenvironment and regulate traction forces \cite{Bischofs2009,Ghibaudo2009,Fletcher2010}, with prominent consequences on bio-mechanical processes such as cell division, differentiation, growth, death, nuclear deformation and gene expression \cite{Minc2011,Versaevel2012,Chen1997,Jain2013,McBeath2004,Kilian2010}. On the other hand, the cellular shape itself depends on the mechanical properties of the environment. Experiments on adherent cells have shown that the stiffness of the substrate strongly affects cell morphology \cite{Chopra2011,Engler2004} and triggers the formation of stress fibers \cite{Yeung2005,Grinnell2000}. The cell spreading area increases with the substrate stiffness for several cell types, including cardiac myocytes \cite{Chopra2011}, myoblasts \cite{Engler2004}, endothelial cells and fibroblasts \cite{Yeung2005}, and mesenchymal stem cells \cite{Zemel2010a}. 

In our previous work \cite{Pomp2018} we have investigated the shape and traction forces of adherent cells \cite{Schwarz2013} characterized by a highly anisotropic actin cytoskeleton. Using a contour model of cellular adhesion \cite{Bar-Ziv1999,Bischofs2008,Bischofs2009,Schwarz2013,Giomi2019}, we demonstrated that the edge of these cells can be accurately approximated by a collection of elliptical arcs obtained from a unique ellipse, whose eccentricity depends on the degree of anisotropy of the contractile stresses arising from the actin cytoskeleton.   Furthermore, our model quantitatively predicts how the anisotropy of the actin cytoskeleton determines the magnitudes and directions of traction forces. Both predictions were tested in experiments on highly anisotropic fibroblastoid and epithelioid cells \cite{Danen2002} supported by stiff microfabricated elastomeric pillar arrays \cite{Tan2003,Trichet2012,VanHoorn2014}, finding good quantitative agreement. 

\begin{figure}[t]
\centering
\includegraphics[width=\columnwidth]{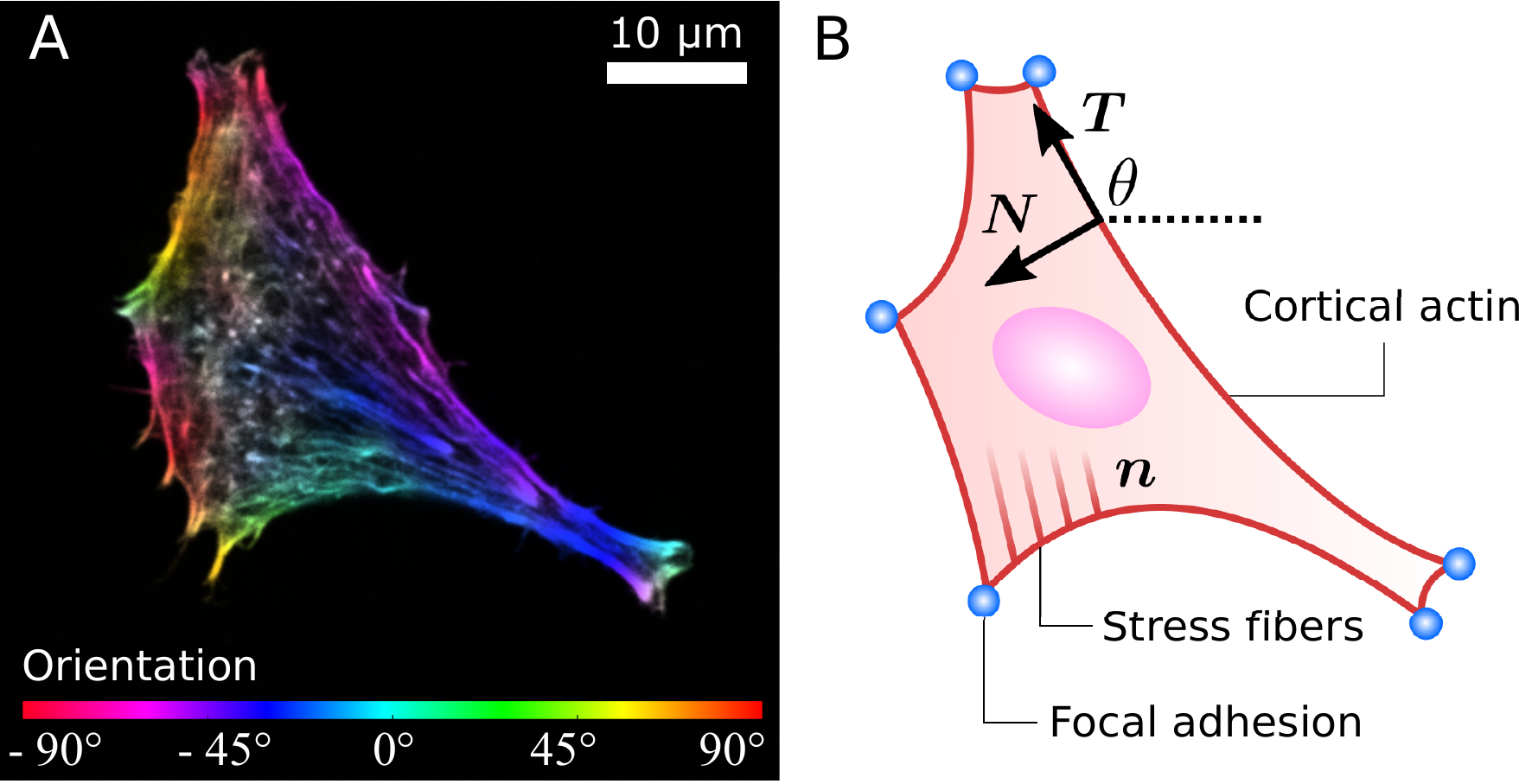}
\caption{\label{fig:schematic} (A) A cell with an anisotropic actin cytoskeleton (fibroblastoid) cultured on a stiff microfabricated elastomeric pillar array \cite{Pomp2018}. The color scale indicates the measured orientation of the actin stress fibers. (B) Schematic representation of a contour model for the cell in (A). The cell contour consists of a collection of concave cellular arcs (red lines) that connect pairs of adhesion sites (blue dots). These arcs are parameterized as curves spanned counterclockwise around the cell by the arclength $s$, and are entirely described via the tangent unit vector $\bm{T}=(\cos\theta,\sin\theta)$ and the normal vector $\bm{N}=(-\sin\theta,\cos\theta)$, with $\theta$ the turning angle. The vector $\bm{n}$ describes the local orientation of the stress fibers.}
\end{figure}

Whereas these findings shed light on how cytoskeletal anisotropy controls the geometry and forces of adherent cells, they raise questions on how anisotropy emerges from the three-fold interplay between isotropic and directed stresses arising within the cytoskeleton, the shape of the cell envelope and the geometrical constraints introduced by focal adhesions. It is well known that the orientation of the stress fibers in elongated cells strongly correlates with the polarization direction of the cell \cite{Vignaud2012,Ladoux2016,Lam2016,Gupta2019}. Consistently, our results indicate that, in highly anisotropic cells, stress fibers align with each other and with the cell's longitudinal direction (see, e.g., Fig. \ref{fig:schematic}A) \cite{Pomp2018}. However, the physical origin of these alignment mechanisms is less clear and inevitably leads to a chicken-and-egg causality dilemma: do the stress fibers align along the cell's axis or does the cell elongate in the direction of the stress fibers?

The biophysical literature is not scarce of cellular processes that might possibly contribute to alignment of stress fibers with each other and with the cell edge. Mechanical feedback between the cell and the extracellular matrix or substrate, for instance, has been shown to play an important role in the orientation and alignment of stress fibers \cite{Deshpande2007,Walcott2010,Zemel2010a,Zemel2010b,Nisenholz2014}. Molecular motors have also been demonstrated to produce an aligning effect on cytoskeletal filaments, both in mesenchymal stem cells \cite{Raab2012} and in purified cytoskeletal extracts \cite{Schaller2010}, where the observation is further corroborated by agent-based simulations \cite{Kraikivski2006}. A similar mechanism has been theoretically proposed for microtubules-kinesin mixtures \cite{Swaminathan2009}. Studies in microchambers demonstrated that steric interactions can also drive alignment of actin filaments with each other and with the microchamber walls \cite{Soares2011,Alvarado2014,Deshpande2012}. Theoretical studies have highlighted the importance of the stress fibers' assembly and dissociation dynamics \cite{Deshpande2006,Deshpande2007}, the dynamics of focal adhesion complexes \cite{Deshpande2008,Rens2018}, or both \cite{Pathak2008,Ronan2014}. Also the geometry of actin nucleation sites has been shown to affect the growth direction of actin filaments, thus promoting alignment \cite{Reymann2010,Letort2015}. Finally, mechanical coupling between the actin cytoskeleton and the plasma membrane has been theoretically shown to lead to fiber alignment, as bending moments arising in the membrane result into torques that reduce the amount of splay within the filaments \cite{Liu2008}. Despite such a wealth of possible mechanisms, it is presently unclear which one or which combination is ultimately responsible for the observed alignment of stress fibers between each other and with the cell's longitudinal direction. Analogously, it is unclear to what extent these mechanisms are sensitive to the specific mechanical and topographic properties of the environment, although some mechanisms rely on specific environmental conditions that are evidently absent in certain circumstances (e.g., the mechanical feedback between the cell and the substrate discussed in Refs. \cite{Thomopoulos2005,Deshpande2007,Zemel2010b} relies on deformations of the substrate and is unlikely to play an important role in experiments performed on stiff micro-pillar arrays).

In this paper we investigate the interplay between the anisotropy of the actin cytoskeleton and the shape of cells adhering to stiff microfabricated elastomeric pillar arrays \cite{Tan2003,Trichet2012,VanHoorn2014}. Rather than pinpointing a specific alignment mechanism, among those reviewed above, we focus on the \emph{interplay} between cell shape and the spatial organization of the actin cytoskeleton. This is achieved by means of a phenomenological treatment of the stress fiber orientation based on the continuum description of nematic liquid crystals, coupled with a contour model of the cell edge \cite{Pomp2018}. The paper is organized as follows: in Sec. \ref{sec_shape} we review in detail our contour model of cells with an anisotropic cytoskeleton \cite{Pomp2018}. We study how the anisotropy affects the curvature of and tension in the cell cortex, as well as the forces that the cell exerts on the substrate. In Sec. \ref{sec_interplay} we further generalize this approach by studying the mechanical interplay between the orientation of the actin cytoskeleton, modeled as a nematic liquid crystal under confinement of the cell edge, and the shape of the cell, and we compare our results to experimental data on highly anisotropic cells. In both sections we assume that the coordinates of the adhesion sites are constant in time and known. A theoretical description of the dynamics of these adhesion sites, as a result of focal adhesion dynamics, is beyond the scope of this study and can be found, for example, in Refs. \cite{Deshpande2008,Rens2018}.

\section{Equilibrium configuration of the cell boundary} 
\label{sec_shape}

Many animal cells spread out after coming into contact with a stiff adhesive surface. They develop transmembrane adhesion receptors at a limited number of adhesion sites that lie mainly along the cell contour (i.e., focal adhesions \cite{Burridge1996}). These cells are then essentially flat and assume a typical shape characterized by arcs which span between the sites of adhesion, while forces are mainly contractile \cite{Schwarz2013}. This makes it possible to describe adherent cells as two-dimensional contractile films, whose shape is unambiguously identified by the position $\bm{r}=(x,y)$ of the cell contour \cite{Bar-Ziv1999,Bischofs2008,Bischofs2009,Banerjee2013,Giomi2013,Pomp2018,Giomi2019}. Fig. \ref{fig:schematic}B illustrates a schematic representation of the cell (fibroblastoid) in Fig. \ref{fig:schematic}A, where the cell contour consists of a collection of curves, referred to as ``cellular arcs", that connect two consecutive adhesion sites. These arcs are parameterized by the arclength $s$ as curves spanned counterclockwise around the cell, 
oriented along the tangent unit vector $\bm{T} = \partial_{s}\bm{r} = (\cos \theta, \sin\theta)$, with $\theta=\theta(s)$ the turning angle of the arc with respect to the horizontal axis of the frame of reference. The normal vector $\bm{N} = \partial_{s}\bm{r}^{\perp} = (-\sin\theta, \cos \theta)$, with $\bm{r}^{\perp} = (-y,x)$, is directed toward the interior of the cell. The equation describing the shape of a cellular arc is obtained upon balancing all the conservative and dissipative forces experienced by the cell contour. These are:
\begin{equation}\label{LP}
\xi_t \partial_{t}\bm{r} =\; \partial_{s}\bm{F}_{\rm cortex}+(\bm{\hat\Sigma}_{\rm out}-\bm{\hat\Sigma}_{\rm in}) \cdot \bm{N}\;,
\end{equation} 
where $t$ is time, $\xi_t$ is a (translational) friction coefficient, $\bm{\hat\Sigma}_{\rm out}$ and $\bm{\hat\Sigma}_{\rm in}$ are the stress tensors on the two sides of the cell boundary and $\bm{F}_{\rm cortex}$ is the stress resultant along the cell contour \cite{Schwarz2013,Bischofs2008,Bischofs2009,Banerjee2013,Pomp2018,Giomi2019}. The temporal evolution of the cell contour is then dictated by a competition between internal and external bulk stresses acting on the cell boundary and the contractile forces arising within the cell cortex. We assume the dynamics of the cell contour to be overdamped.

The stress tensor can be modeled upon taking into account isotropic and directed stresses. The latter are constructed by treating the stress fibers as contractile force dipoles, whose average orientation $\theta_{\rm SF}$ is parallel to the unit vector $\bm{n}=(\cos\theta_{\rm SF},\sin\theta_{\rm SF})$ (see Fig. \ref{fig:schematic}B). This gives rise to an overall contractile bulk stress of the form \cite{Pedley1992,Simha2002}:
\begin{equation}
\label{eq_bulkstress}
\bm{\hat\Sigma}_{\rm out}-\bm{\hat\Sigma}_{\rm in}=\sigma\bm{\hat I} +\alpha\bm{n}\bm{n}\;,
\end{equation}
where $\bm{\hat I}$ is the identity matrix, $\sigma>0$ embodies the magnitude of all isotropic stresses (passive and active) experienced by the cell edge and $\alpha>0$ is the magnitude of the directed contractile stresses and is proportional to the local degree of alignment between the stress fibers, in such a way that $\alpha$ is maximal for perfectly aligned fibers, and vanishes if these are randomly oriented. In Sec. \ref{sec_interplay} we will explicitly account for the local orientational order of the stress fibers using the language of nematic liquid crystals. The degree of anisotropy of the bulk stress is thus determined by the ratio between the isotropic contractility $\sigma$ and the directed contractility $\alpha$. Finally, the tension within the cell cortex is modeled as $\bm{F}_{\rm cortex}=\lambda \bm{T}$, where line tension $\lambda$ embodies the contractile forces arising from myosin activity in the cell cortex. This quantity varies, in general, along an arc and can be expressed as a function of the arclength $s$. In the presence of anisotropic bulk stresses, in particular, $\lambda(s)$ cannot be constant, as we will see in Sec. \ref{sec_equilibrium}. The force balance condition, Eq. \eqref{LP}, becomes then
\begin{equation}
\label{shape}
\xi_t \partial_{t}\bm{r}\; =\; \partial_{s}(\lambda\bm{T})+\sigma\bm{N}+\alpha(\bm{n}\cdot\bm{N})\bm{n}\;.
\end{equation}

\subsection{Equilibrium cell shape and line tension}

\label{sec_equilibrium}
In this section we describe the position of the cell boundary under the assumption that the timescale required for the equilibration of the forces in Eq. \eqref{shape} is much shorter than the typical timescale of cell migration on the substrate (i.e., minutes). Under this assumption, $\partial_t \bm{r} = \bm{0}$ and Eq. \eqref{shape} can be cast in the form:
\begin{equation}
\label{force_balance2}
\bm{0}\; =\; (\partial_{s}\lambda)\,\bm{T}+(\lambda\kappa+\sigma)\bm{N}+\alpha(\bm{n}\cdot\bm{N})\bm{n}\;,
\end{equation}
where we have used $\partial_{s}\bm{T} = \kappa \bm{N}$, with $\kappa=\partial_{s}\theta$ the curvature of the cell edge. A special situation is obtained when the cytoskeleton is purely isotropic. In this case, first discussed by Bar-Ziv {\em et al}. in the context of cell pearling \cite{Bar-Ziv1999} and later expanded by Bischofs and coworkers \cite{Bischofs2008,Bischofs2009}, $\alpha=0$ and Eq. \eqref{force_balance2} reduces to
\begin{equation}
\label{isotropic_shape}
(\partial_{s}\lambda)\,\bm{T}+(\lambda\kappa+\sigma)\bm{N} =\bm{0}\;.
\end{equation}
Solving this equation in the two orthogonal directions $\bm{T}$ and $\bm{N}$ leads to two important insights for isotropic cells. First, in the presence of strictly isotropic bulk forces, the line tension $\lambda$ must be constant along a single arc at equilibrium (i.e., $\partial_{s}\lambda=0$). Second, bulk and peripheral contractility, that pull the cell edge inward and outward along the normal direction, compromise along an arc of constant curvature $\kappa=-\sigma/\lambda$, with the negative sign of $\kappa$ indicating that the arcs are curved inwards. This mechanism is analogous to the Young-Laplace law for fluid interfaces, where the isotropic bulk contractility $\sigma$ plays the role of the pressure difference and the line tension $\lambda$ that of surface tension. The cell edge is then described by a sequence of circular arcs, whose radius $R=1/|\kappa|=\lambda/\sigma$ depends on the local cortical tension $\lambda$ of the arc. The latter depends on the local myosin concentration and, in general, varies from arc to arc. The possibility of a geometry-dependent cortical tension, originating from the elasticity of the transverse fibers, was also explored in Refs. \cite{Bischofs2008,Bischofs2009}, to account for an observed correlation between the radius and the length of the cellular arcs. Both geometry-dependent and geometry-independent models successfully describe the shape of adherent cells in the presence of strictly isotropic forces. However, as we showed elsewhere \cite{Pomp2018}, these isotropic models are not suited for describing cells that develop strong directed forces due to the anisotropic cytoskeleton originating from actin stress fibers \cite{Pellegrin2007,Burridge2013}. 

The presence of an anisotropic cytoskeleton is modeled by a nonzero directed stress, $\alpha>0$. By virtue of Eq. \eqref{force_balance2}, this results into the fact that the cortical tension $\lambda$ is no longer constant along the cellular arcs, since the directed stress has, in general, a non-vanishing tangential component (i.e., $\bm{n}\cdot\bm{T} \ne 0$). In order to highlight the physical mechanisms described by Eq. \eqref{force_balance2}, we introduce a number of simplifications in the remainder of this section. These will be lifted in Sec. \ref{sec_interplay}, where we will consider the most general scenario. First, because the orientation of the stress fibers typically varies only slightly along a single arc \cite{Pomp2018}, we assume the orientation of the stress fibers, $\theta_{\rm SF}$, to be constant along a single cellular arc, but different from arc to arc. Furthermore, we assume that the contractile bulk stresses $\alpha$ and $\sigma$ are uniform throughout the cell. This allows us to solve Eq. \eqref{force_balance2} analytically for a given arc. Without loss of generality, we orient the reference frame such that the stress fibers are parallel to the $y-$axis. Thus, $\theta_{\rm SF}=\pi/2$ and $\bm{n} =\bm{\hat{y}}$. Since $\alpha$, $\sigma$ and $\bm{n}$ are constant along an arc, Eq. \eqref{force_balance2} can be expressed as a total derivative and integrated directly. This yields
\begin{equation}
\label{eq:integrated}
\lambda\bm{T}+(\sigma\bm{\hat{I}}+\alpha\bm{n}\bm{n})\cdot\bm{r}^{\perp} = \bm{C}_{1}\;,
\end{equation}
where $\bm{C}_{1}=(C_{1x},C_{1y})$ is an integration constant. Decomposing Eq. (\ref{eq:integrated}) into $x-$ and $y-$directions yields
\begin{subequations}\label{scalar_force_balance}
\begin{align}
\lambda\cos\theta &= C_{1x} +\sigma y\, \label{scalar_force_balance_a} \\ 
\lambda\sin\theta &= C_{1y} - (\alpha+\sigma)x\;. \label{scalar_force_balance_b}
\end{align}
\end{subequations}
Next, taking the ratio of Eqs. \eqref{scalar_force_balance}, using $\tan\theta=\dd y/\dd x$ and integrating, we obtain a general solution for the shape of the cellular arc subject to a non-vanishing isotropic stress (i.e., $\sigma \ne 0$), namely
\begin{equation}\label{eq:ellipse1}
\frac{1}{\gamma}(x -x_c)^2 +(y -y_c)^{2} = C_{2}\;,
\end{equation}
where $C_{2}$ is another integration constant and we have set
\begin{equation*}
x_{c} = \frac{C_{1y}}{\sigma+\alpha}\;,\qquad
y_{c} =-\frac{C_{1x}}{\sigma}\;,\qquad
\gamma = \frac{\sigma}{\sigma+\alpha}\;.	
\end{equation*}	
Eq. \eqref{eq:ellipse1} describes an ellipse centered at $(x_{c},y_{c})$ and whose minor and major semi-axis are $a=\sqrt{\gamma C_{2}}$ and $b=\sqrt{C_{2}}$. The constant $\gamma$ quantifies the anisotropy of the bulk stress: $\gamma = 1$ corresponds to the isotropic case, whereas $\gamma=0$ represents purely anisotropic bulk contractility. Using again Eqs. \eqref{scalar_force_balance}, we further obtain an expression for the line tension $\lambda$ as a function of $x$ and $y$:
\begin{equation}
\label{eq:tension1}
\lambda^2 = \sigma^2 (y -y_c)^2 + (\sigma +\alpha)^2 (x -x_c)^2\;.
\end{equation}
Using Eqs. \eqref{scalar_force_balance} and \eqref{eq:ellipse1}, this can be also expressed as a function of the turning angle $\theta$, namely
\begin{equation}
\label{eq:tension2}
\frac{\lambda^{2}}{\sigma^{2}} = C_{2}\,\frac{1+\tan^{2}\theta}{1+\gamma\tan^{2}\theta}\;.
\end{equation}
This expression highlights the physical meaning of the integration constant $C_{2}$. The right-hand side of Eq. \eqref{eq:tension2} attains its minimal value ($C_2$) where $\theta=0$, hence when the tangent vector is perpendicular to the stress fibers (i.e., $\bm{n}\cdot\bm{T}=0$). Thus $C_{2}=\lambda_{\min}^{2}/\sigma^{2}$, where $\lambda_{\rm min}$ is the minimal tension withstood by the cortical actin. Substituting $C_2$ in Eq. \eqref{eq:tension2} yields the tension as a function of the turning angle:
\begin{equation}\label{eq_tension}
\lambda(\theta) = \lambda_{\mathrm{min}} \sqrt{\frac{1 + \tan^{2}\theta}{1 + \gamma \tan^{2}\theta}}\;.
\end{equation}
Eq. \eqref{eq_tension} describes how the line tension deviates from the minimum value $\lambda_{\min}$ for nonzero $\theta$ in order to balance the tangential component of the directed stress [Eq. \eqref{force_balance2}]. The maximum value of the line tension is found at $\theta = \pi/2$, where the stress fibers are parallel to the arc, and is given by $\lambda_{\max}=\lambda_{\min}/\sqrt{\gamma}$. We note that these variations in line tension along a single arc do not necessarily have to be regulated by the cell. Instead, they could simply be a result of passive mechanical forces in a way very similar to the space-dependent tension in a simple cable hanging under gravity.

\begin{figure}[t]
\centering
\includegraphics[width=0.9\columnwidth]{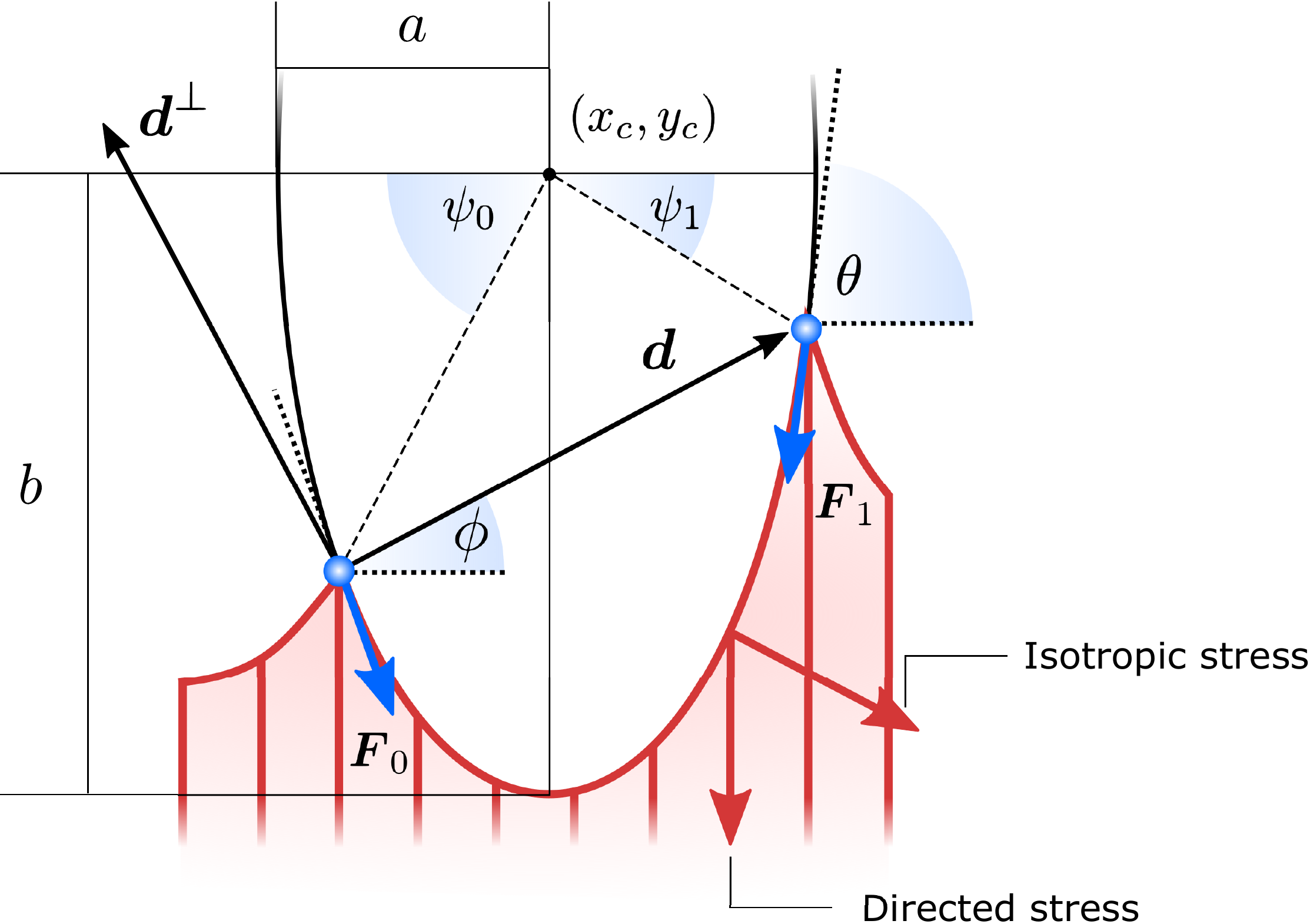}
\caption{\label{fig2} Schematic representation of a cellular arc, described by Eq. (\ref{eq_ellipse}), for $\theta_{\mathrm{SF}} = \pi/2$. A force balance between isotropic stress, directed stress and line tension results in the description of each cell edge segment (red curve) as part of an ellipse of aspect ratio $a/b=\sqrt\gamma$. The cell exerts forces $\bm F_0$ and $\bm F_1$ on the adhesion sites (blue). The vector $\bm{d} = d (\cos\phi,\sin\phi)$ describes the relative position of the two adhesion sites, ${\bm{d}}^\perp= d(-\sin\phi,\cos\phi)$ is a vector perpendicular to $\bm{d} $, and $\theta$ is the turning angle of the cellular arc. The coordinates of the ellipse center $(x_{c},y_{c})$ and the angular coordinates of the adhesion sites along the ellipse $\psi_0$ and $\psi_1$ are given in Appendix A.}
\end{figure}

Although the minimal line tension $\lambda_{\mathrm{min}}$ could, in principle, be arc-dependent, for example if the cell cortex displays substantial variations in the myosin densities \cite{Bischofs2008}, here we approximate $\lambda_{\rm min}$ as a constant. This approximation is motivated by the fact that our previous {\em in vitro} observations of anisotropic epithelioid and fibroblastoid cells did not identify a correlation between the arc length and curvature \cite{Pomp2018}, which, on the other hand, is expected if $\lambda_{\mathrm{min}}$ was to vary significantly from arc to arc \cite{Bischofs2008}. Hence, $\alpha$, $\sigma$ and $\lambda_{\rm \min}$ represent the material parameters of our model.

Substituting $C_2$ in Eq. \eqref{eq:ellipse1} yields an implicit representation of the plane curve approximating individual cellular arcs, namely
\begin{equation}\label{eq:ellipse_final}
\frac{\sigma^{2}}{\gamma\lambda_{\min}^{2}}\, (x -x_c)^{2} +\frac{\sigma^{2}}{\lambda_{\min}^{2}}\, (y - y_c)^{2} = 1\;.
\end{equation}
This equation describes an ellipse centered at the point $(x_{c},y_{c})$ and oriented along the $y-$direction, whose minor and major semi-axes are $a=\lambda_{\min}\sqrt{\gamma}/\sigma$ and $b=\lambda_{\min}/\sigma$ respectively (Fig. \ref{fig2}). For arbitrary stress fiber orientation $\theta_{\rm SF}$, Eq. \eqref{eq:ellipse_final} can be straightforwardly generalized in the form:
\begin{multline}\label{eq_ellipse}
\frac{\sigma^{2}}{\lambda_{\min}^{2}}\left[(x-x_{c})\cos \theta_{\rm SF}+(y-y_{c})\sin\theta_{\rm SF}\right]^{2}\\
+ \frac{\sigma^{2}}{\gamma\lambda_{\min}^{2}}\left[-(x-x_{c})\sin \theta_{\rm SF}+(y-y_{c})\cos\theta_{\rm SF}\right]^{2} = 1\;.
\end{multline}
Consistently, the major axis of the ellipse is always parallel to the stress fibers, hence tilted by an angle $\theta_{\rm SF}$ with respect to the $x-$axis (Fig. \ref{fig2}). The direct relation between the contractile forces arising from the cytoskeleton and the shape of the cell is highlighted by the dimensionless parameter $\gamma = \sigma/(\sigma +\alpha)$: on the one hand, $\gamma$ defines the anisotropy of the contractile bulk stress,  on the other hand it dictates the anisotropy of the cell shape. The coordinates of the center of the ellipse $(x_{c},y_{c})$ and the angular coordinates of the adhesion sites along the ellipse, $\psi_0$ and $\psi_1$ in Fig. \ref{fig2}, can be calculated using standard algebraic manipulation and are given in Appendix A.

\begin{figure}[t]
\centering
\includegraphics[width=1\columnwidth]{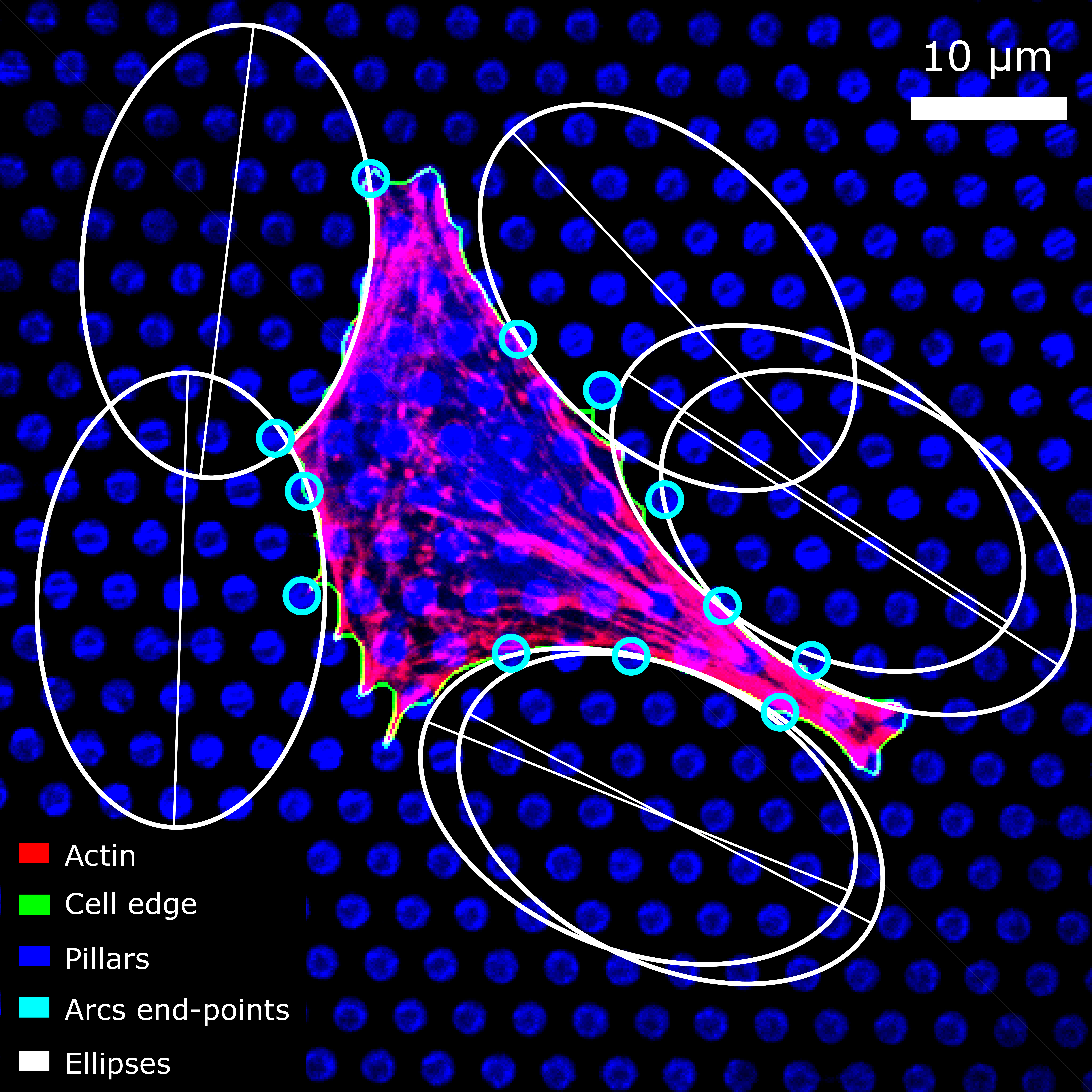}
\caption{\label{fig3} A cell with an anisotropic actin cytoskeleton on a microfabricated elastomeric pillar array (fibroblastoid \cite{Pomp2018}, same cell as in Figure \ref{fig:schematic}A), with a unique ellipse (white) fitted to its arcs. The actin, cell edge, and micropillar tops are in the red, green, and blue channels respectively. The endpoints of the arcs (cyan) are identified based on the forces that the cell exerts on the pillars. For a detailed description of the experimental methods, see Ref. \cite{Pomp2018}. Scale bar is 10 $\mathrm{\mu m}$.}
\end{figure}

One of the most striking consequences of the elliptical shape of the cellular arcs is that the local curvature is no longer constant, consistent with experimental observations on epithelioid and fibroblastoid cells in Ref. \cite{Pomp2018}. This can be calculated from Eq. \eqref{eq:ellipse_final} using $\kappa=(\dd^2 y/\dd x^{2})/[1+(\dd y/\dd x)^{2}]^{\frac{3}{2}}$ and expressed in terms of the turning angle $\theta$ using Eqs. \eqref{scalar_force_balance} and \eqref{eq_tension}. This yields
\begin{equation}
\label{eq_kappax}
\kappa 
=-\frac{1}{\gamma b} \left(\frac{1+\gamma \: \tan^{2}\theta}{1+\tan^{2}\theta}\right)^{\frac{3}{2}}\;,
\end{equation}
with the negative sign indicating that the arcs are curved inwards. A cellular arc thus attains its maximal (minimal) absolute curvature, where $\theta=0$ ($\theta=\pi/2$) and the stress fibers are parallel (perpendicular) to the arc tangent vector, namely 
\begin{subequations}
\begin{align}
\kappa_{\min} &= \kappa\left(\theta=\frac{\pi}{2}\right) = -\frac{\sqrt{\gamma}}{b}\;,\\
\kappa_{\max} &= \kappa\left(\theta=0\right) = -\frac{1}{\gamma b}\;.
\end{align}
\end{subequations}
Consistent with experimental evidence \cite{Pomp2018}, the radius of curvature of arcs perpendicular to stress fibers is smaller than the radius of curvature of arcs parallel to the stress fiber direction. For the experimental methods and a more detailed comparison between theory and experiment, see Ref. \cite{Pomp2018}. We stress that, as long as the contractile stresses arising from the actin cytoskeleton are roughly uniform across the cell (i.e., $\alpha$, $\sigma$ and $\lambda_{\mathrm{min}}$ are constant), all cellular arcs are approximated by a {\em unique} ellipse (see, e.g., Fig. \ref{fig3}). Thus, regions of the cell edge having higher and lower local curvature correspond to different portions of the same ellipse, depending on the relative orientation of the local tangent vector and the stress fibers.

\subsection{Traction forces}

With the expressions for shape of the cellular arcs [Eq. \eqref{eq_ellipse}] and cortical tension [Eq. \eqref{eq_tension}] in hand, we can now calculate the traction forces exerted by the cell via the focal adhesions positioned at the end-points of a given cellular arc (Fig. \ref{fig2}). Calling these $\bm{F}_{0}$ and $\bm{F}_{1}$ and recalling the the cell edge is oriented counter-clockwise, we have $\bm{F}_{0}=-\lambda(\theta_{0})\bm{T}(\theta_{0})$ and $\bm{F}_{1}=\lambda(\theta_{1})\bm{T}(\theta_{1})$, where $\theta_{0}$ and $\theta_{1}$ are the turning angles at the end-points of the arc. For practical applications, it is often convenient to express the position of the adhesion sites in terms of their relative distance $\bm{d}=d(\cos\phi,\sin\phi)$ (Fig. \ref{fig2}). This yields
\begin{subequations}\label{eq_forces}
\begin{multline}\label{eq_traction_forces}
\bm{F}_{0} 
= \lambda_{\mathrm{min}}\Biggr[\Biggr(\frac{d}{2b}\sin \phi + \frac{\rho}{b}  \cos \phi \Biggr) \: \bm{n}^{\perp} \\ 
+ \Biggr(-\frac{1}{\gamma} \frac{d}{2b} \cos \phi  + \frac{\rho}{b} \sin \phi \Biggr) \; \bm{n} \, \Biggr],
\end{multline}
\begin{multline}
\bm{F}_{1} 
= \lambda_{\mathrm{min}}\Biggr[\Biggr(\frac{d}{2b}\sin \phi - \frac{\rho}{b}  \cos \phi \Biggr) \: \bm{n}^{\perp} \\ 
+ \Biggr(-\frac{1}{\gamma} \frac{d}{2b} \cos \phi  - \frac{\rho}{b} \sin \phi \Biggr) \; \bm{n} \, \Biggr],
\end{multline}
\end{subequations}
where $\bm{n}^{\perp} = (\sin \theta_{\mathrm{SF}}, -\cos \theta_{\mathrm{SF}})$ and the length scale $\rho$ is defined as
\begin{equation}
\label{eq_rho}
\rho = \sqrt{b^{2} \left(\frac{1 +\tan^{2} \phi}{1 + \gamma \tan^{2} \phi}\right) - \frac{1}{\gamma}\left(\frac{d}{2}\right)^{2}}\;.
\end{equation}
The total traction force exerted by the cell can be calculated by summing the two forces associated with the arcs joining at a given adhesion site, while taking into account that the the orientation $\bm{n}$ of the stress fibers is generally different from arc to arc.

Another interesting quantity is obtained by adding the forces $\bm{F}_0 $ and $\bm{F}_1$ from the same arc. Although these two forces act on two different adhesion sites, their sum represents the total net force that a single cellular arc exerts on the substrate. This is given by
\begin{align}
\label{eq_netforce3}
\bm{F}_0 + \bm{F}_1 
&= d \sigma \sin \phi\,\bm{n}^{\perp} - d (\sigma + \alpha) \cos \phi \,\bm{n}\;,\notag\\
&= -\left(\sigma \bm{\hat{I}} + \alpha\,\bm{n}\bm{n}\right) \cdot \bm{d}^{\perp}\;,
\end{align}
where $\bm{d}^{\perp}=d(-\sin\phi,\cos\phi)$ (Fig. \ref{fig2}). Eq. \eqref{eq_netforce3} is the force resulting from the integrated contractile bulk stress [see Eq. \eqref{LP}] and is independent of the line tension $\lambda_{\min}$. 

\subsection{Mechanical invariants}

We conclude this section by highlighting two mechanical invariants, local quantities that are constant along a cellular arc, thus showing the intimate relation between the geometry of the cell and the mechanical forces it exerts on the environment. From Eqs. \eqref{eq_forces} we find
\begin{equation}
\label{eq:constant1}
F_{\perp}^2 + \gamma F_{\parallel}^2 = \mathrm{const.},
\end{equation}
where $F_{\parallel}$ and $F_{\perp}$ are the components of the force, parallel and perpendicular to $\bm{n}$, at any point along a same cellular arc. Furthermore, by inspection of Eqs. \eqref{eq_kappax} and \eqref{eq_tension} we observe that
\begin{equation}
\label{eq:constant2}
\lambda^3 \kappa = -\lambda_{\mathrm{min}}^2 (\alpha+\sigma) = {\rm const}\;.
\end{equation}
From this, we find that the normal component of the cortical force, $\lambda \kappa$ [see Eq. \eqref{force_balance2}], is then given by
\begin{equation}
\label{eq:constant3}
\lambda \kappa = -\left(\frac{\lambda_{\mathrm{min}}}{\lambda}\right)^2 (\alpha +\sigma)\;.
\end{equation}
This relation is an analog of the Young-Laplace law for our anisotropic system. In the isotropic limit, $\alpha=0$ and $\lambda_{\min}=\lambda$, thus we recover the standard force-balance expression $\lambda \kappa = -\sigma$. Eq. \eqref{eq:constant3} shows that the normal force $\lambda \kappa$ decreases with increasing line tension $\lambda$, because an increase in line tension is accompanied by an even stronger decrease in the curvature $\kappa$.

\section{Interplay between orientation of the cytoskeleton and cellular shape}
\label{sec_interplay}
In this Section we generalize our approach by allowing the orientation of the stress fibers to vary along individual cellular arcs. This is achieved by combining the contour model for the cell shape, reviewed in Sec. \ref{sec_shape}, with a continuous phenomenological model of the actin cytoskeleton, rooted into the hydrodynamics of nematic liquid crystals \cite{Gennes_book}. This setting can account for the mechanical feedback between the orientation of the stress fibers and the cellular shape and allows us to predict both these features starting from the position of the adhesions sites alone. A treatment of the dynamics of focal adhesions is beyond the scope of this paper and can be found elsewhere, e.g., in Refs. \cite{Deshpande2008,Rens2018}.

As mentioned in the Introduction, experimental observations, by us \cite{Pomp2018} and others \cite{Vignaud2012,Ladoux2016,Lam2016,Gupta2019}, have indicated that stress fibers tend to align with each other and with the cell's longitudinal direction. As we discussed, several cellular processes might contribute to these alignment mechanisms, such as mechanical cell-matrix feedback \cite{Zemel2010a,Zemel2010b,Deshpande2007,Walcott2010,Nisenholz2014}, motor-mediated alignment \cite{Raab2012,Schaller2010,Kraikivski2006,Swaminathan2009}, steric interactions \cite{Soares2011,Alvarado2014,Deshpande2012}, stress fiber formation and dissociation \cite{Deshpande2007,Deshpande2006,Pathak2008,Ronan2014}, focal adhesion dynamics \cite{Deshpande2008,Rens2018,Pathak2008,Ronan2014}, the geometry of actin nucleation sites \cite{Reymann2010,Letort2015}, or membrane-mediated alignment \cite{Liu2008}, but it is presently unclear which combination of mechanisms is ultimately responsible for the orientational correlation observed in experiments. Our phenomenological description of the actin cytoskeleton allows us to focus on the \emph{interplay} between cellular shape and the orientation of the stress fibers, without the loss of generality that would inevitably result from selecting a specific alignment mechanism among those listed above. 

\subsection{Dynamics of the stress fibers}
\label{sec_bulk}

The actin cytoskeleton is modeled as a nematic liquid crystal confined within the cellular contour. This is conveniently represented in terms of the two-dimensional nematic tensor:
\begin{equation}\label{eq:q_tensor}
Q_{ij} = S \left(n_{i}n_j-\frac{1}{2}\,\delta_{ij}\right)\;,	
\end{equation}
where $\delta_{ij}$ is the Kronecker delta and $S=\sqrt{2\tr\bm{Q}^{2}}$ is the so called nematic order parameter, measuring the amount of local nematic order. In the context of the actin cytoskeleton, a low $S$ value might result from either a low local density of stress fibers, or from a high density of randomly oriented stress fibers. In the standard $\{\bm{\hat{x}},\bm{\hat{y}}\}$ Cartesian basis, Eq. \eqref{eq:q_tensor} yields
\begin{equation}
\label{eq_Q_tensor}
\bm{\hat{Q}} =
\begin{bmatrix}
 Q_{xx}& Q_{xy} \\ 
 Q_{xy}& -Q_{xx}
\end{bmatrix}
= \frac{S}{2}
\begin{bmatrix}
 \cos{2\theta_{\mathrm{SF}}}& \sin{2\theta_{\mathrm{SF}}}\\ 
 \sin{2\theta_{\mathrm{SF}}}&  -\cos{2\theta_{\mathrm{SF}}}
\end{bmatrix}.
\end{equation}
Just like the dynamics of the cell edge, the dynamics of the cytoskeleton are assumed to be overdamped, thus:
\begin{equation}\label{eq:eom1}
\partial_{t}Q_{ij} = -\frac{1}{\xi_{r}}\,\frac{\delta F_{\rm cyto}}{\delta Q_{ij}}\;,
\end{equation}
where $\xi_{r}$ is a rotational friction coefficient, dictating the rate of the relaxational dynamics, and $F_{\rm cyto}$ is the Landau-de Gennes free-energy \cite{Gennes_book}: 
\begin{multline}
\label{eq:energy}
F_{{\rm cyto}} 
= \frac{K}{2} \int \dd A\, \left[ |\nabla \bm{\hat{Q}}|^2 + \frac{1}{\delta^2} \tr \bm{\hat{Q}}^2 (\tr\bm{\hat{Q}}^2 - 1) \right] \\ 
+ \frac{W}{2} \oint \dd s \tr \left[(\bm{\hat{Q}} - \bm{\hat{Q}}_{0})^2\right]\;.
\end{multline}
The first integral in Eq. \eqref{eq:energy} corresponds to a standard mean-field free-energy, favoring perfect nematic order (i.e., $S=1$), while penalizing gradients in the orientation of the stress fibers and their local alignment. $K$ is the orientational stiffness of the nematic phase in the one elastic constant approximation. The length-scale $\delta$ sets the size of the boundary layer in regions where the order parameter drops to zero to compensate a strong distortion of the nematic director $\bm{n}$, such as in proximity of topological defects. The second integral, which is extended over the cell contour, is the Nobili-Durand anchoring energy \cite{Nobili1992} and determines the orientation of the stress fibers along the edge of the cell, with the tensor $\bm{\hat{Q}}_{0}$ representing their preferential orientation. Experimental evidence form our (Fig. \ref{fig3} and Ref. \cite{Pomp2018}) and other's work (e.g., Refs. \cite{Vignaud2012,Ladoux2016,Lam2016,Gupta2019}), suggests that, in highly anisotropic cells, peripheral stress fibers are preferentially parallel to the cell edge. The same trend is recovered in experiments with purified actin bundles confined in microchambers \cite{Soares2011,Alvarado2014}. In the language of Landau-de Gennes theory, this effect can be straightforwardly reproduced by setting
\begin{equation}
\label{eq_q0}
Q_{0,ij} =  T_i T_j -\frac{1}{2}\, \delta_{ij}\;,	
\end{equation}
with $\bm{T}$ the tangent unit vector of the cell edge, under the additional assumption that nematic order is prominent along the cell contour (i.e., $S =1$). The constant $W>0$ measures the strength of this parallel anchoring. To ensure the free-energy to be minimal at equilibrium (i.e., when $\partial_t Q_{ij} = 0$), we solve Eq. \eqref{eq:eom1} with Neumann boundary conditions:
\begin{equation}\label{eq:bc}
K \bm{N}\cdot \nabla Q_{ij} - 2 W \left(Q_{ij} - Q_{0,ij}\right) = 0\;.
\end{equation}

\subsection{The dynamics of the cell edge}
\label{sec_edge_dynamics}
The relaxational dynamics of the cell contour are governed, in our model, by Eq. \eqref{shape}, which is now lifted from the assumption that the orientation $\bm{n}$ of the stress fibers is uniform along individual cellular arcs. Furthermore, the contractile stress given by Eq. \eqref{eq_bulkstress} can now be generalized as:
\begin{equation}\label{eq_gen_bulkstress}
\bm{\hat{\Sigma}}_{\rm out}-\bm{\hat{\Sigma}}_{\rm in} = \left(\sigma+\frac{1}{2}\,\alpha_{0}S\right)\bm{\hat{I}} + \alpha_{0}\bm{\hat{Q}}\;,
\end{equation}
with $\alpha_{0}$ a constant independent on the local order parameter. Comparing Eqs. \eqref{eq_bulkstress} and \eqref{eq_gen_bulkstress} yields $\alpha=\alpha_{0}S$, thus the formalism introduced in this Section allows us to explicitly account for the effect of the local orientational order on the amount of contractile stress exerted by the stress fibers.

Next, we decompose Eq. \eqref{shape} along the normal and tangent directions of the cell contour. Since the cells considered here are pinned at the adhesion sites and the density of actin along the cell contour is assumed to be constant, tangential motion is suppressed, i.e., $\bm{T}\cdot\partial_{t}\bm{r}=0$. Together with Eq. \eqref{eq_gen_bulkstress} this yields:
\begin{subequations}
\label{eq:decomposed}
\begin{align}
0  &= \partial_{s}\lambda\,+\alpha_0 \, \bm{T} \cdot \bm{\hat{Q}} \cdot \bm{N} \;, \label{eq:decomposed_a} \\[5pt] 
\xi_t \bm{N}\cdot\partial_{t}\bm{r} &= \lambda\kappa+\sigma\,+\frac{1}{2}\,\alpha_0 S +\alpha_0\, \bm{N} \cdot \bm{\hat{Q}} \cdot \bm{N} \;. \label{eq:decomposed_b}
\end{align}
\end{subequations}
Eq. (\ref{eq:decomposed_a}) describes then the relaxation of tension $\lambda$ within the cell edge, given its shape, whereas Eq. \eqref{eq:decomposed_b} describes the relaxation of the cellular shape itself. The variations in the cortical tension might result from a regulation of the myosin activity or simply form a passive response of the cortical actin to the tangential stresses.

Integrating Eq. \eqref{eq:decomposed_a} then yields the cortical tension along an arc:
\begin{equation}
\label{eq:lambda}
\lambda(s) = \lambda(0) - \alpha_0 \int_0^s\dd s'\, \bm{T} \cdot \bm{\hat{Q}} \cdot \bm{N} \;,
\end{equation}
where $\bm{\hat{Q}}=\bm{\hat{Q}}(s)$ varies, in general, along an individual cellular arc. The integration constant $\lambda(0)$, which represents the cortical tension at one of the adhesion sites, is related to the minimal tension $\lambda_{\min}$ withstood by the cortical actin which we used, in Sec. \ref{sec_shape}, as material parameter of the problem. In practice, we first calculate $\lambda$ over an entire arc using a arbitrary guess for $\lambda(0)$. Then, we apply a uniform shift in such a way that the minimal $\lambda$ value coincides with $\lambda_{\min}$.

Combining the dynamics of the cell contour and that of the cell bulk, we obtain the following coupled differential equations:
\begin{subequations}\label{eq:eom2}
\begin{align}
\partial_{t}\bm{r} &= \frac{1}{\xi_{t}}\,\left[\lambda\kappa+\sigma+\frac{1}{2}\,\alpha_0 S +\alpha_0\, \bm{N} \cdot \bm{\hat{Q}} \cdot \bm{N}\right] \bm{N} \, , \label{eq:eom2a} \\[5pt] 
\partial_{t}\bm{\hat{Q}} &= \frac{K}{\xi_{r}}\,\left[\nabla^{2}\bm{\hat{Q}}-\frac{2}{\delta^{2}}\,(S^{2}-1)\bm{\hat{Q}}\right]\;. \label{eq:eom2b}
\end{align}	
\end{subequations}
These are complemented by the condition that $\bm{r}$ is fixed in a number of specific adhesion sites, the boundary condition given by Eq. \eqref{eq:bc} for the nematic tensor $\bm{\hat{Q}}$ and the requirement that $\min_{s}\lambda(s)=\lambda_{\min}$ on each arc.

\section{Numerical results}
\label{results}

Eqs. \eqref{eq:eom2} are numerically solved using a finite difference integration scheme with moving boundary. As we detail in Appendix \ref{numerical_scheme}, the time-integration is performed iteratively using the forward Euler algorithm by alternating updates of the cell contour and of the bulk nematic tensor. This process is iterated until both the cell shape and the orientation reach a steady state. 

To highlight the physical meaning of our numerical results, we introduce two dimensionless numbers, namely the {\em contractility} number, {\em Co}, and the {\em anchoring} number, {\em An}. {\em Co} is defined as the ratio between the typical distance between two adhesion sites $d$ and the major axis of the ellipse approximating the corresponding cellular arc ($b =\lambda_{\mathrm{min}}/\sigma $, see Sec. \ref{sec_equilibrium}):
\begin{equation}
\label{contractility}
\text{\em Co} = \frac{\sigma d}{\lambda_{\mathrm{min}}}\;,
\end{equation}
and provides a dimensionless measure of the cell contraction (thus of the cell average curvature). The anchoring number, on the other hand, is defined as the ratio between a typical length scale $R$ in which the internal cell structure is confined (not necessarily equal to the distance $d$) and the length scale $K/W$, which sets the size of the boundary layer where $\bm{\hat{Q}}$ crosses over from its bulk configuration to $\bm{\hat{Q}}_{0}$:
\begin{equation}
\label{anchoring}
\text{\em An} = \frac{WR}{K}\;. 
\end{equation}
This number expresses the ratio between the anchoring energy, which scales as $WR$ [i.e., last term in Eq. (\ref{eq:energy})], and the bulk energy, which scales as $K$, thus independently on cell size [i.e., first term in Eq. (\ref{eq:energy})]. Hence, {\em An} represents the competition between boundary alignment (with strength $W$) and bulk alignment (strength $K$) within the length scale of the cell $R$. For $\text{\em An} \ll 1$, bulk elasticity dominates over boundary anchoring and the orientation of the stress fibers in the bulk propagates into the boundary, resulting into a uniform orientation throughout the cell and large deviations from parallel anchoring in proximity of the edge. Conversely, for $\text{\em An} \gg 1$, boundary anchoring dominates bulk elasticity and the orientation of the stress fibers along the cell edge propagates into the bulk, leading to non-uniform alignment in the interior of the cell.

To get insight on the effect of the combinations of these dimensionless parameters on the spatial organization of the cell, we first consider the simple case in which the adhesion sites are located at the corners of squares and rectangles (Sec. \ref{rectangular}). In Sec. \ref{sec_real} we consider more realistic adhesion geometries and compare our numerical results with experimental observations on highly anisotropic cells.

\subsection{Rectangular cells}
\label{rectangular}

\begin{figure}[t!]
\centering
\includegraphics[width=\columnwidth]{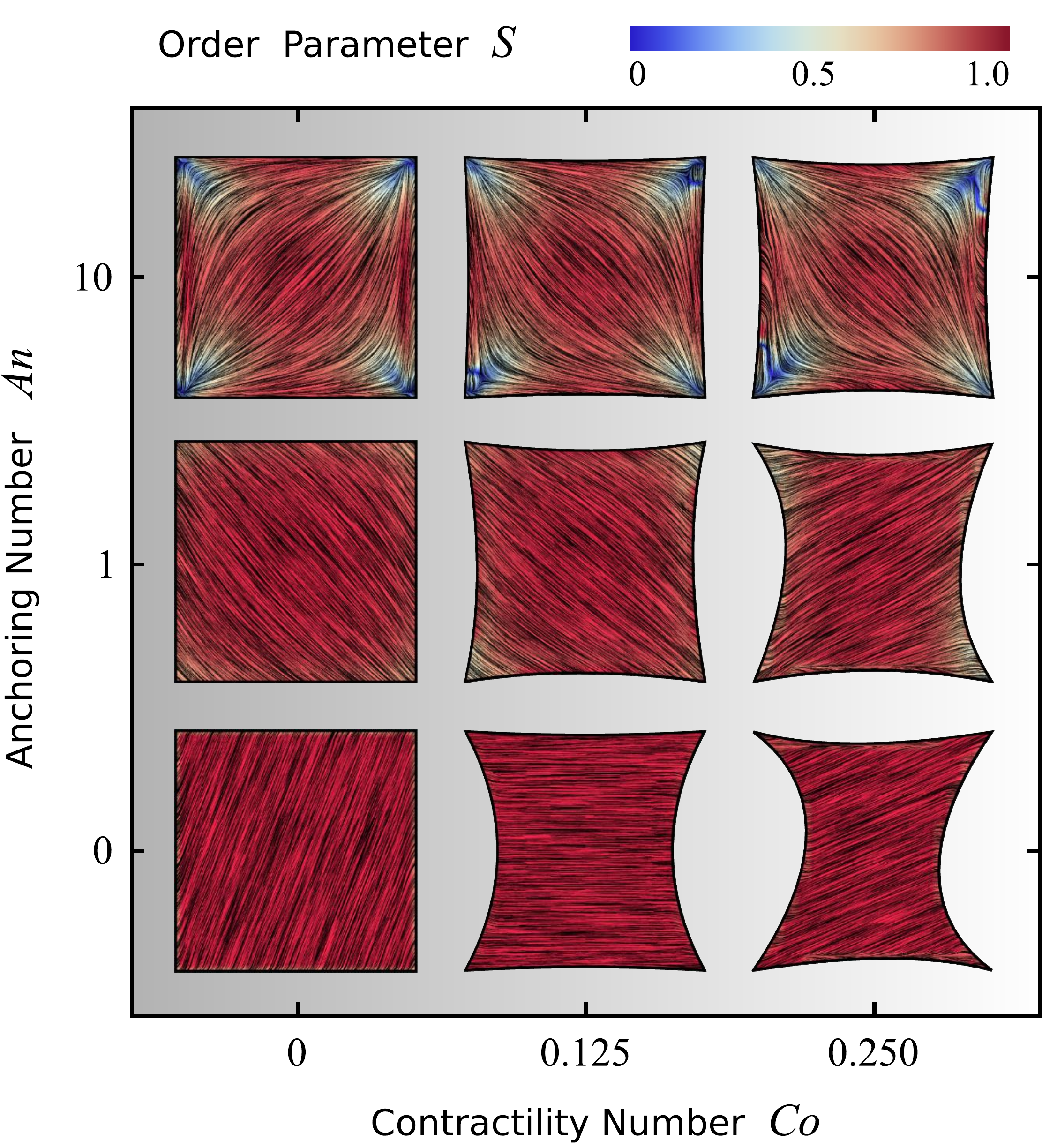}
\caption{Configurations of cells whose adhesion sites are located at the vertices of a square. The thick black line represents the cell boundary, the black lines in the interior of the cells represent the orientation field $\bm{n}=(\cos\theta_{\mathrm{SF}},\sin \theta_{\mathrm{SF}})$ of the stress fibers and the background color indicates the local nematic order parameter $S$. The vertical axis corresponds to the anchoring number $\text{\em An} =WR/K$ and the horizontal axis to the contractility number $\text{\em Co} = \sigma d/\lambda_{\min}$. The cells shown correspond to values of $\text{\em An} = 0, 1, 10$ and $\text{\em Co} = 0, 0.125, 0.25$, where we take both $d$ and $R$ equal to the length of the square side. The ratios $\sigma/(\sigma +\alpha_0)=1/9$, $\lambda_{\mathrm{min}}\Delta t/(\xi_t R^2) = 2.8 \cdot 10^{-6}$,  and $K\Delta t/(\xi_r R^2) = 2.8 \cdot 10^{-6}$, and the parameters $\delta = 0.15R$, $N_\mathrm{arc} = 20$, and $\Delta x = R/19$ are the same for all cells. The number of iterations is $5.5\cdot 10^5$. For definitions of $\Delta t$, $\Delta x$, and $N_\mathrm{arc}$, see Appendix \ref{numerical_scheme}.}
\label{fig:phase_diagram}
\end{figure}

Fig. \ref{fig:phase_diagram} shows the possible configurations of a model cell whose adhesion sites are located at the vertices of a square, obtained upon varying {\em An} and {\em Co}, while keeping $\gamma = \sigma/(\sigma +\alpha_0)$ constant. The thick black line represents the cell boundary, the black lines in the interior of the cells represent the orientation field $\bm{n}$ of the stress fibers and the background color indicates the local nematic order parameter $S$.

As expected, for low {\em Co} values (left column), cells with high {\em An} exhibit better parallel anchoring than cells with small {\em An} values, but lower nematic order parameter in the cell interior. Interestingly, the alignment of stress fibers with the walls in the configuration with large {\em An} value (top left) resembles the configurations found by Deshpande \emph{et al.} \cite{Deshpande2006,Deshpande2007}, who specifically accounted for the assembly and dissociation dynamics of the stress fibers. More strikingly, the structure reported in the top left of Fig. \ref{fig:phase_diagram} appears very similar to those found in experiments of actin filaments in cell-sized square microchambers \cite{Soares2011,Alvarado2014}, simulations of hard rods in quasi-2D confinement \cite{Soares2011}, and results based on Frank elasticity \cite{Galanis2006}, where the tendency of the nematic director to align along the square edges competes with that of aligning along the diagonal. 

For large {\em Co} values (right column of Fig. \ref{fig:phase_diagram}), the cell deviates from the square shape. Interestingly, {\em An} and {\em Co} do not influence the geometry of the cell independently. For constant {\em Co}, i.e., for fixed stress fiber contractility, increasing {\em An} leads to higher tangential alignment of the stress fibers with the cell edge, thus increasing {\em An} decreases the contractile force experienced by the cell edge, which is proportional to $(\bm{n}\cdot\bm{N})^{2}$ [Eq. (\ref{eq:eom2a})].

Finally, we note that all configurations in Fig. \ref{fig:phase_diagram} display a broken rotational and/or chiral symmetry. For $\text{\em An} =0$ the stress fibers are uniformly oriented, but any direction is equally likely. For non-zero {\em An}, the stress fibers tend to align along either of the diagonals (with the same probability) to reduce the amount of distortion. Upon increasing {\em Co}, chirality emerges in the cytoskeleton and in the cell contour (see, e.g., the cell in the middle of the right column in Fig. \ref{fig:phase_diagram}). In light of the recent interest in chiral symmetry breaking in single cells \cite{Tee2015} and in multicellular environments \cite{Duclos2018}, we find it particularly interesting that chiral symmetry breaking can originate from the natural interplay between the orientation of the stress fibers and the shape of the cell.  

\begin{figure}[t]
\centering
\includegraphics[width=\columnwidth]{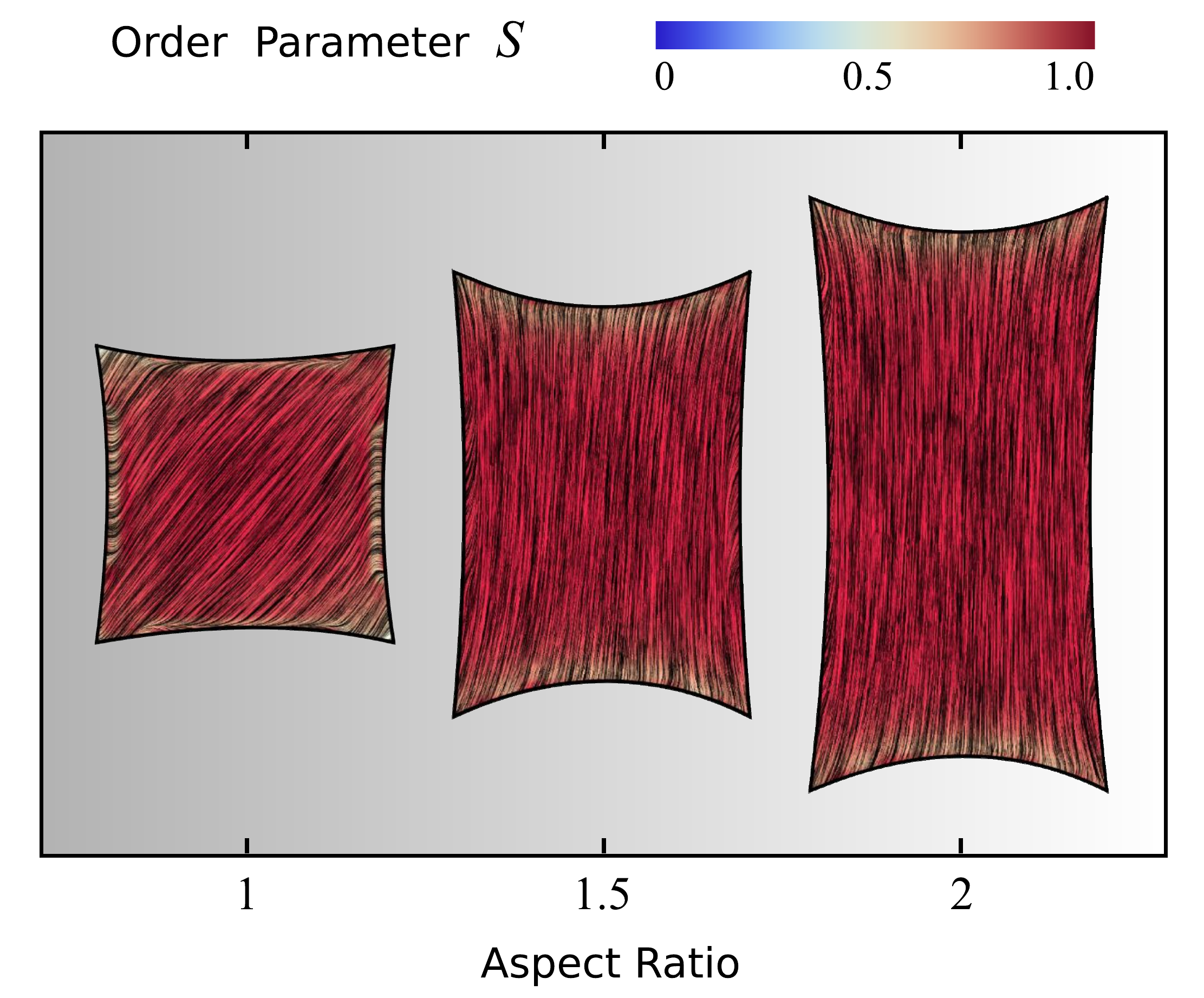}
\caption{Effect of the aspect ratio, ranging from 1 to 2, of the cell on cytoskeletal organization for cells whose four adhesion sites are located at the vertices of a rectangle. The thick black line represents the cell boundary, the black lines in the interior of the cells represent the orientation field $\bm{n}=(\cos\theta_{\mathrm{SF}},\sin \theta_{\mathrm{SF}})$ of the stress fibers and the background color indicates the local nematic order parameter $S$. The simulations shown are performed with $\text{\em An} = WR/K =1$ where $R$ is equal to the short side of the rectangle, and $\text{\em Co} = \sigma d/\lambda_{\mathrm{min}}$ equal to 0.125, 0.1875, and 0.25 respectively, where $d$ is equal to the long side of the rectangle. The ratios $\sigma/(\sigma +\alpha_0)=1/9$, $\lambda_{\mathrm{min}}\Delta t/(\xi_t R^2) = 2.8 \cdot 10^{-6}$,  and $K\Delta t/(\xi_r R^2) = 2.8 \cdot 10^{-6}$, and the parameters $\delta = 0.15R$ and $\Delta x = R/19$ are the same for all cells. $N_\mathrm{arc} = 20, 30, 40$ from left to right and the number of iterations is $5.5\cdot 10^5$. For definitions of $\Delta t$, $\Delta x$, and $N_\mathrm{arc}$, see Appendix \ref{numerical_scheme}.}
\label{fig_aspect_ratio}
\end{figure}

To conclude this section, we focus on four-sided cells whose adhesion sites are located at the vertices of a rectangle and explore the effect of the cell aspect ratio (i.e., height/width). Fig. \ref{fig_aspect_ratio} displays three configurations with aspect ratio increasing from 1 to 2. Upon increasing the cell aspect ratio, the mean orientation of the stress fibers switches from the diagonal (aspect ratio 1) to longitudinal (aspect ratio 2), along with an increase in the order parameter in the cell bulk, as can be seen in Fig. \ref{fig_aspect_ratio} by the slightly more red-shifted cell interior. This behavior originates from the competition between bulk and boundary effects. Whereas the bulk energy favors longitudinal alignment, as this reduces the amount of bending of the nematic director, the anchoring energy favors alignment along all four edges alike, thus favoring highly bent configurations at the expense of the bulk elastic energy. When the aspect ratio increases, the bending energy of the bulk in the diagonal configuration increases, whereas the longitudinal state only pays the anchoring energy at the short edges, hence independently on the aspect ratio. Therefore, elongating the cell causes the stress fibers to transition from tangential to longitudinal alignment, with a consequent increase of the nematic order parameter. Interestingly, these observations are consistent with the findings of experiments on actin filaments in cell-sized microchambers \cite{Soares2011,Alvarado2014}.

\subsection{Cells on micropillar arrays}
\label{sec_real}

\begin{figure}[t!]
\centering
\includegraphics[width=1\columnwidth]{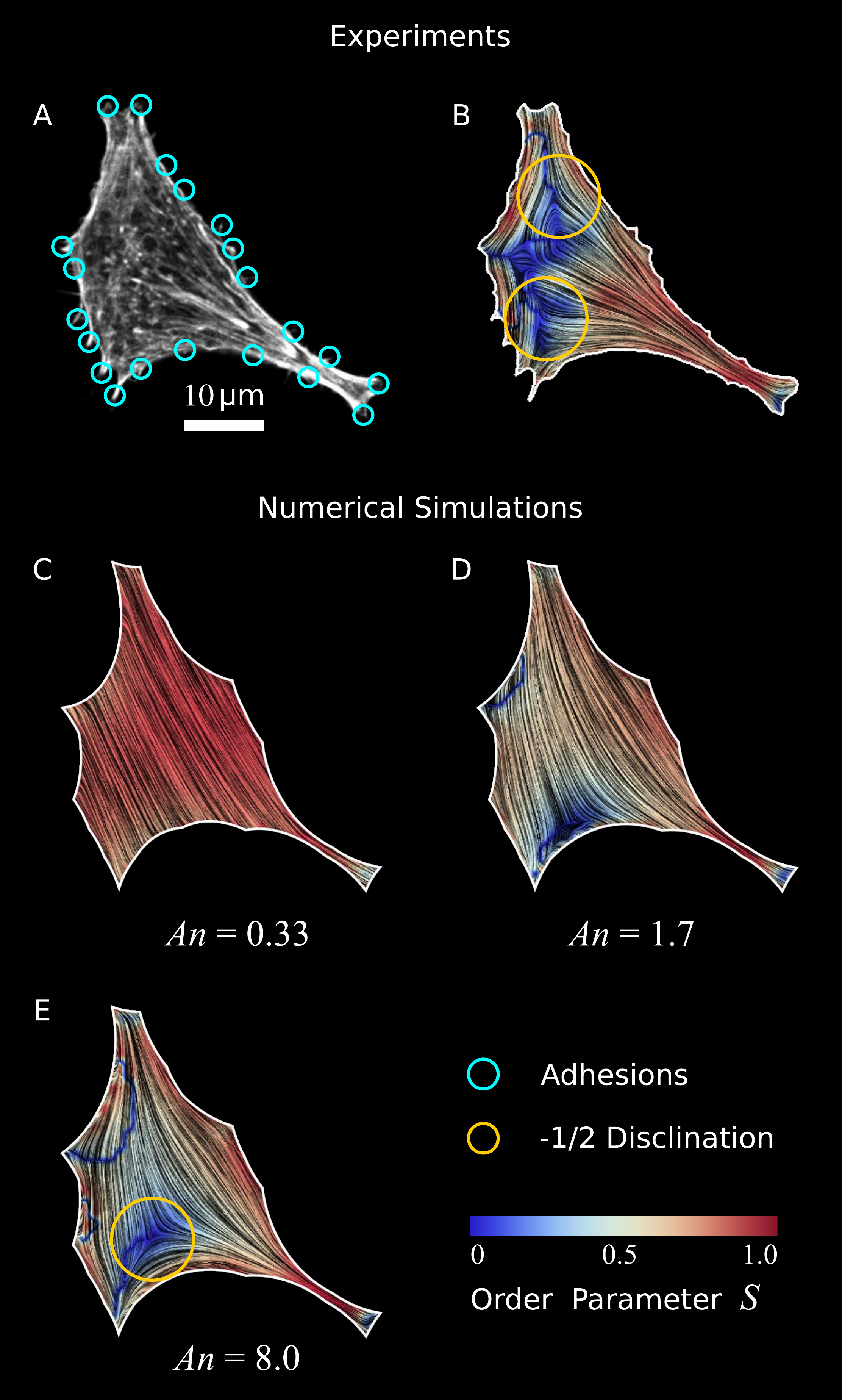}
\caption{Validation of our model to experimental data. (A) Optical micrograph of a fibroblastoid cell (same cell as in Figs. \ref{fig:schematic} and \ref{fig3}) \cite{Pomp2018}. (B) Experimental data of cell shape and coarse-grained cytoskeletal structure of this cell. The white line represents the cell boundary, black lines in the interior of the cells represent the orientation field $\bm{n}=(\cos\theta_{\mathrm{SF}},\sin \theta_{\mathrm{SF}})$ of the stress fibers and the background color indicates the local nematic order parameter $S$. (C-E) Configurations obtained from a numerical solution of Eqs. \eqref{eq:eom2} using the adhesion sites (cyan circles) of the experimental data as input and with various anchoring number ({\em An}) values. This is calculated from Eq. \eqref{anchoring}, with $R = 23.6\:\mu $m, taken from the square root of the cell area. The corresponding values of the length scale $K/W$ are $71\:\mu $m (C), $14\:\mu $m (D) and $2.9\:\mu $m (E) respectively. The values for $\lambda_{\mathrm{min}}/\sigma = 14.7 \: \mu$m and $\sigma/(\sigma +\alpha_0) = 0.40$ are found by an analysis of the elliptical shape of this cell \cite{Pomp2018}. The ratios $\lambda_{\mathrm{min}}\Delta t/\xi_t = 1.2 \cdot 10^{-3} \mu\mathrm{m}^2$  and $K\Delta t/\xi_r = 1.2\cdot 10^{-3} \mu\mathrm{m}^2$, and the parameters $\delta = 11\:\mu$m, $N_\mathrm{arc} = 20$, and $\Delta x = 1.1\:\mu$m are the same for figures (C-E). The number of iterations is $2.1\cdot 10^6$. For definitions of $\Delta t$, $\Delta x$, and $N_\mathrm{arc}$, see Appendix \ref{numerical_scheme}.\\}
\label{fig_cell_example}
\end{figure}

In order to validate our model experimentally, we compare our numerical results with experiments on fibroblastoid and epithelioid cells \cite{Danen2002} plated on stiff micropillar arrays \cite{Tan2003,Trichet2012,VanHoorn2014}. The cells are imaged using spinning disk confocal microscopy (see, e.g., Fig. \ref{fig_cell_example}A) and the images are then processed in order to detect the orientation of the stress fibers. Upon coarse-graining the local gradients of the image intensity, we reconstruct both the nematic director $\bm{n}$ (black lines, representing the orientation of the stress fibers) and order parameter $S$ (background color, representing the degree of alignment), as visualized in Fig. \ref{fig_cell_example}B. See Appendix \ref{experimental_methods} for more detail on the experimental methods and image processing. 

Consistent with our results on rectangular cells (Fig. \ref{fig_aspect_ratio}), the stress fibers align parallel to the cell's longitudinal direction and perpendicularly to the cell's shorter edges. Furthermore, the nematic order parameter is close to unity in proximity of the cell contour, indicating strong orientational order near the cell edge, but drops in the interior. This behavior is in part originating from the lower density of stress fibers around the center of mass of the cell, and in part from the presence of $\pm 1/2$ nematic disclinations away from the cell edge. These topological defects naturally arise from the tangential orientation along the boundary. Albeit not uniform throughout the whole cell contour, thus not sufficient to impose hard topological constraints on the configuration of the director in the bulk (i.e., Poincar\'e-Hopf theorem), this forces a non-zero winding of the stress fibers, which in turn is accommodated via the formation of one or more disinclinations. As a consequence of the concave shape of the cell boundary, these defects have most commonly strength $-1/2$.

To compare our theoretical and experimental results, we extract the locations of the adhesion sites from the experimental data (see Ref. \cite{Pomp2018}) and use them as input parameters for the simulations. In Figs. \ref{fig_cell_example}C-E we show results of simulations of the cell in Figs. \ref{fig_cell_example}A,B for increasing {\em An} values, thus decreasing magnitude of the length scale $K/W$. Here, we take the length scale $R= 23.6\:\mu$m to be the square root of the cell area and we use constant values for the ratios $\lambda_{\mathrm{min}}/\sigma = 14.7 \: \mu$m and $\sigma/(\sigma +\alpha_0)= 0.40$ as found by an analysis of the elliptical shape of this cell \cite{Pomp2018}. Fig. \ref{fig_cell_example}C shows the results of a simulation where bulk alignment dominates over boundary alignment ({\em An} = 0.33, $K/W = 71\:\mu$m), resulting in an approximately uniform cytoskeleton oriented along the cell's longitudinal direction. The nematic order parameter is also approximatively uniform and close to unity. For larger {\em An} values (Fig. \ref{fig_cell_example}D, {\em An} = 1.7 and $K/W = 14\:\mu $m), anchoring effects become more prominent, resulting in distortions of the bulk nematic director and the emergence of a $-1/2$ disclination in the bottom left side of the cell. Upon further increasing {\em An} (Fig. \ref{fig_cell_example}E, {\em An} = 8.0 and $K/W = 2.9\:\mu $m), the $-1/2$ topological defect moves towards the interior, as a consequence of the increased nematic order along the boundary. This results in a decrease in nematic order parameter in the bulk of the cell, consistent with our experimental data. 

A qualitative comparison between our in {\em vitro} (Fig. \ref{fig_cell_example}B) and {\em in silico} cells (Fig. \ref{fig_cell_example}E), highlights a number of striking similarities, such as the overall structure of the nematic director, the configuration of the order parameter (large along the cell edge and in the thin neck at the bottom right of the cell) and the occurrence of a $-1/2$ disclination on the bottom-left side. In order to make the comparison quantitative and infer the value of the phenomenological parameters introduced in this Section, we have further analyzed the residual function

\begin{figure}
\centering
\includegraphics[width=\columnwidth]{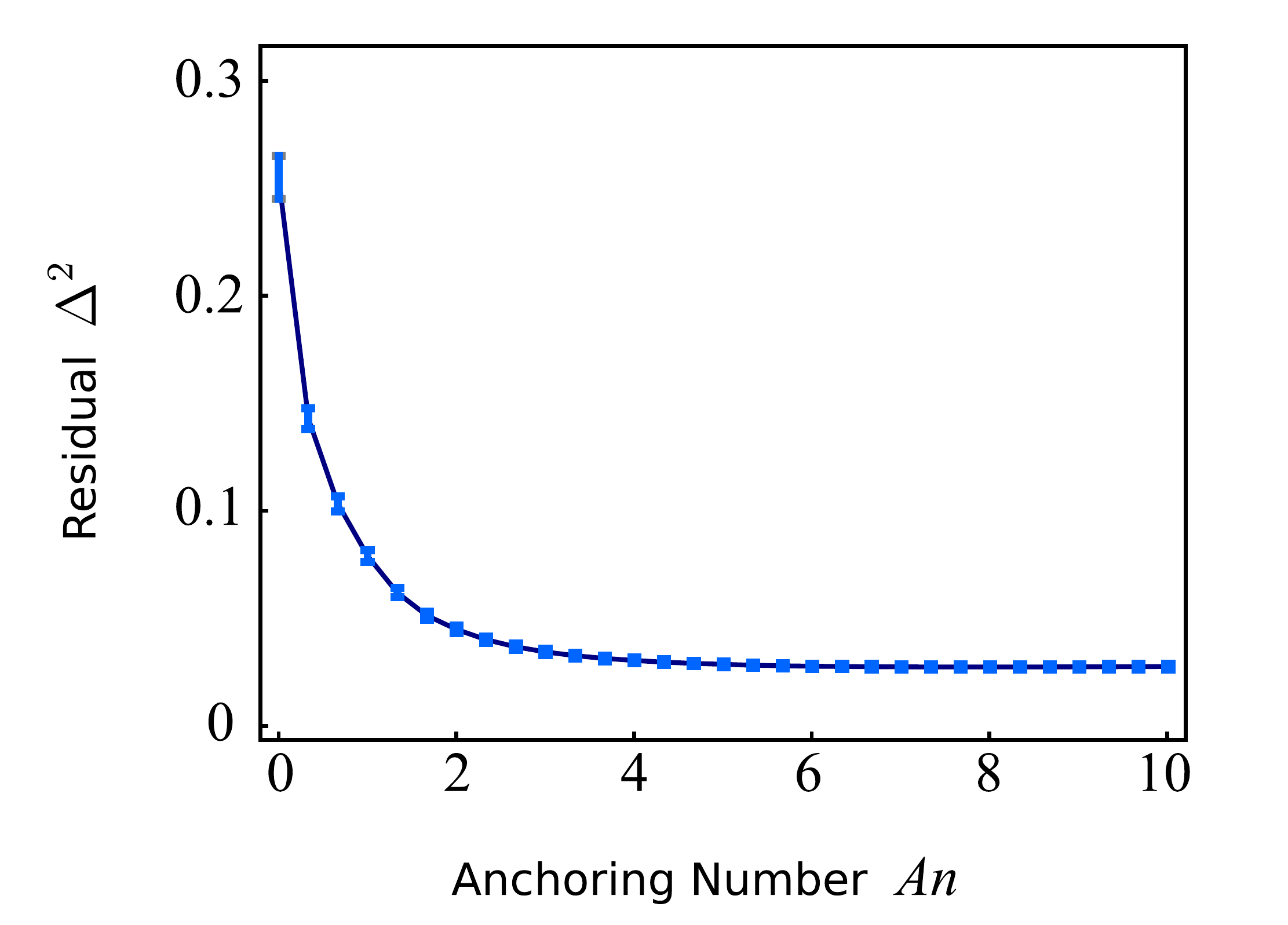}
\caption{Residual function $\Delta^{2}$, defined in Eq. \eqref{eq_Delta}, as a function of the anchoring number {\em An} (Eq. \eqref{anchoring} with $R = 23.6\:\mu$m) for the cell displayed in Fig. \ref{fig_cell_example}. The error bars display the standard deviation obtained using jackknife resampling. For large {\em An} values the residual flattens, indicating that the actual value of {\em An} becomes unimportant once the anchoring torques, which determine the tangential alignment of the stress fibers in the cell's periphery, outcompete the bulk elastic torques. The minimum ($\Delta^2 = 0.027$) is found for {\em An} $= 8.0$.}
\label{fig_delta_square}
\end{figure}

\begin{equation}
\label{eq_Delta}
\Delta^2 = \frac{1}{N} \sum_{i =1}^{N} 
\frac{1}{2} \tr \big[( \bm{\hat{Q}}_{{\rm sim},i} - \bm{\hat{Q}}_{{\rm exp},i})^2\big]\;,
\end{equation}
expressing the departure of the predicted configurations of the nematic tensor, $\bm{\hat{Q}}_{\rm sim}$, from the experimental ones, $\bm{\hat{Q}}_{\rm exp}$. The index $i$ identifies a pixel in the cell and $N$ is the total number of pixels common to both numerical and experimental configurations. By construction, $0\le \Delta^2 \le 1$, with 0 representing perfect agreement. Fig. \ref{fig_delta_square} shows a plot of $\Delta^{2}$ versus the anchoring number {\em An} for the cell shown in Fig. \ref{fig_cell_example}. Consistent with the previous qualitative comparison, the agreement is best at large {\em An} values, indicating a substantial preference of the stress fibers for parallel anchoring along the cell edge. For the cell in Fig. \ref{fig_cell_example}, $\Delta^2$ is minimized for $\text{\em An} = 8.0$ ($\Delta^2 = 0.027$), corresponding to a boundary layer $K/W = 2.9\:\mu$m. The corresponding numerically calculated configuration is shown in Fig. \ref{fig_cell_example}E. However, the flattening of $\Delta^{2}$ for large {\em An} values implies that the actual value of {\em An} becomes unimportant once the anchoring torques, which determine the tangential alignment of the stress fibers in the cell's periphery, outcompete the bulk elastic torques. Therefore, we conclude that the cell illustrated in Fig. \ref{fig_cell_example} is best described by $\text{\em An} \gtrsim  5$, corresponding to a boundary layer $K/W \lesssim 5 \,\mu$m.

\begin{figure}[h!]
\centering
\includegraphics[width=1\columnwidth]{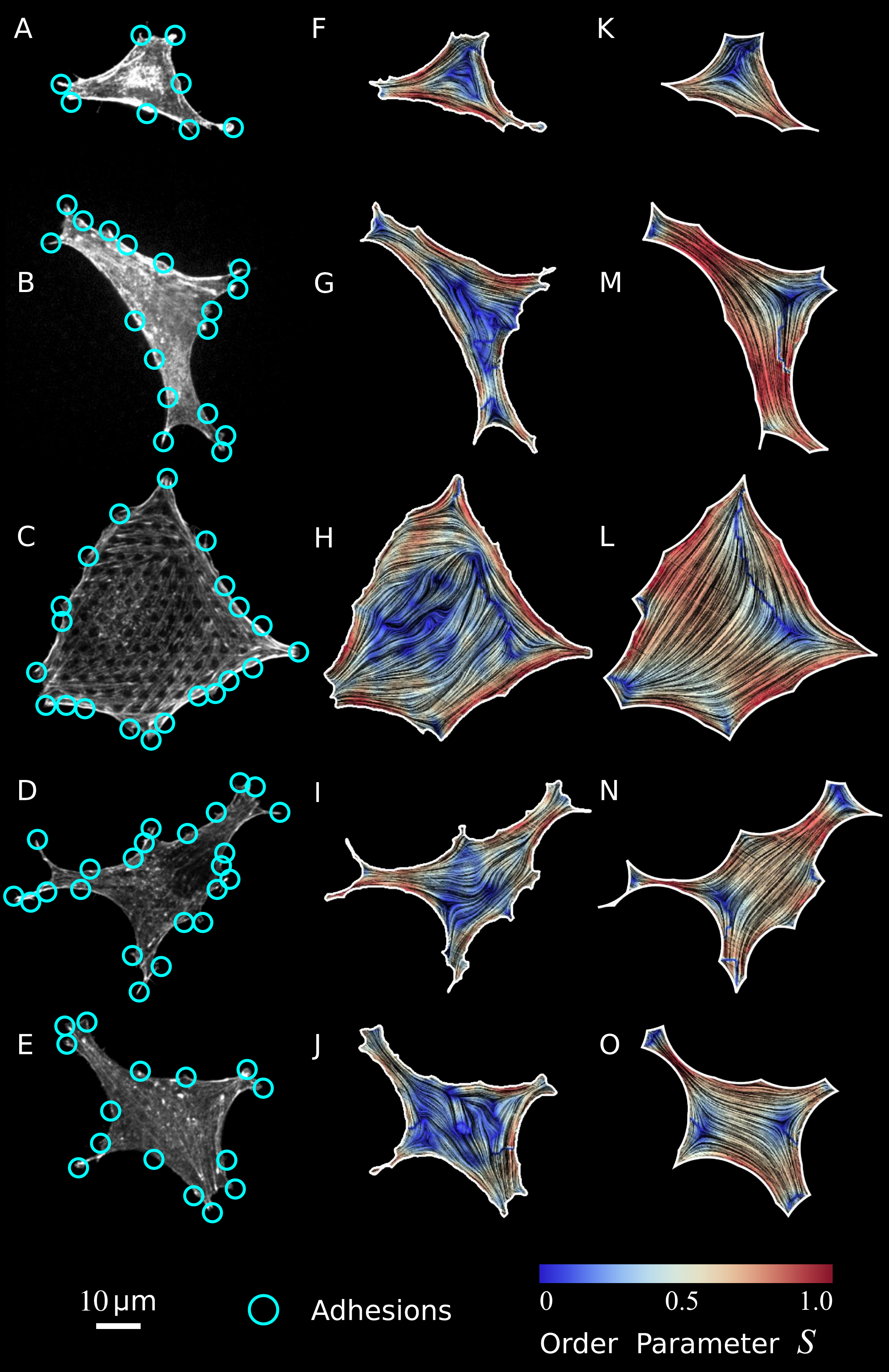}
\caption{Comparison of experimental data on five anisotropic cells with the results of computer simulations. (A-E) Epithelioid (A,B,E) and fibroblastoid (C,D) cells on a microfabricated elastomeric pillar array \cite{Pomp2018}. (F-J) Experimental data of cell shape and coarse-grained cytoskeletal structure on a square lattice of these cells. The white line represents the cell boundary, the black lines in the interior of the cells represent the orientation field $\bm{n}=(\cos\theta_{\mathrm{SF}},\sin \theta_{\mathrm{SF}})$ of the stress fibers and the background color indicates the local nematic order parameter $S$. (K-O) Simulations with the adhesion sites of the experimental data as input. The values for $\lambda_{\mathrm{min}}/\sigma = 12.6; 15.7; 18.0; 10.8; 13.4 \: \mu$m and $\sigma/(\sigma +\alpha_0 ) = 0.75; 0.25; 0.46; 0.95; 0.52 $ are found by an analysis of the elliptical shape of these cells \cite{Pomp2018}. The values of $\text{\em An} = 4.4; 4.1; 19; 4.6; 4.7$, where $R = 17.3; 24.4; 39.9; 24.9; 25.3\:\mu $m is defined as the square root of the cell area, are determined by minimizing $\Delta^2$, with the minima given by $\Delta^2 = 0.016; 0.058; 0.057; 0.034; 0.037$. These {\em An} values correspond to $K/W = 3.9; 5.9; 2.1; 5.4; 5.4\:\mu$m. The ratios $\lambda_{\mathrm{min}}\Delta t/\xi_t = 1.2 \cdot 10^{-3} \mu\mathrm{m}^2$  and $K\Delta t/\xi_r = 1.2\cdot 10^{-3} \mu\mathrm{m}^2$, and the parameters $\delta = 11\:\mu$m, $N_\mathrm{arc} = 20$, and $\Delta x = 1.1\:\mu$m are the same for all cells. The number of iterations is $2.1\cdot 10^6$. For definitions of $\Delta t$, $\Delta x$, and $N_\mathrm{arc}$, see Appendix \ref{numerical_scheme}.\\}
\label{fig_all_cells}
\end{figure}

The same analysis presented above has been repeated for five other cells (Fig. \ref{fig_all_cells}). The first column shows the raw experimental data, the second column shows the coarse-grained experimental data, and the third column shows the simulations. For these we used the values of $\lambda_{\mathrm{min}}/\sigma $ and $\sigma/(\sigma +\alpha_0)$ obtained from a previous analysis of the cell morphology \cite{Pomp2018} and the {\em An} values found by a numerical minimization of $\Delta^{2}$. Also for these cells $\Delta^2$ flattens for large {\em An} values, and we estimate $\text{\em An} \gtrsim 3$ and $K/W \lesssim 7\:\mu$m. Similar to the cell in Fig. \ref{fig_cell_example}, we observe that the overall structure of the nematic director, including the emergence of $-1/2$ topological defects, is captured well by our approach, although there is no precise correspondence between the locations of the defects in the experiments and the theory.

\section{Discussion and conclusions}
\label{sec_discussion}

In this article we investigated the spatial organization of cells adhering on a rigid substrate at a discrete number of points. Our approach is based on a contour model for the cell shape \cite{Bar-Ziv1999,Bischofs2008,Bischofs2009,Schwarz2013,Giomi2019} coupled with a continuous phenomenological model for the actin cytoskeleton, inspired by the physics of nematic liquid crystals \cite{Gennes_book}. This approach can be carried out at various levels of complexity, offering progressively insightful results. Assuming that the orientation of the stress fibers is uniform along individual cellular arcs (but varies from arc to arc), it is possible to achieve a complete analytical description of the geometry of the cell, in which all arcs are approximated by different portions of a unique ellipse. The eccentricity of this ellipse depends on the ratio between the isotropic and directed stresses arising in the actin cytoskeleton, and the orientation of the major axis of this ellipse is parallel to the stress fibers. This method further allows to analytically calculate the traction forces exerted by the cell on the adhesion sites and compare them with traction force microscopy data.

Lifting the assumption that the stress fibers are uniformly oriented along individual cellular arcs allows one to describe the mechanical interplay between cellular shape and the configuration of the actin cytoskeleton. Using numerical simulations and inputs from experiments on fibroblastoid and epithelioid cells plated on stiff micropillar arrays, we identified a feedback mechanism rooted in the competition between the tendency of stress fibers to align uniformly in the bulk of the cell, but tangentially with respect to the cell edge. Our approach enables us to predict both the shape of the cell and the orientation of the stress fibers and can account for the emergence of topological defects and other distinctive morphological features. These predictions are in good agreement with the experimental data and further offer an indirect way to estimate quantities that are generally precluded to direct measurement, such as the cell's internal stresses and the orientational stiffness of the actin cytoskeleton. 

The success of this relatively simple approach is remarkable given the enormous complexity of the cytoskeleton and the many physical, chemical, and biological mechanisms associated with stress fiber dynamics and alignment \cite{Zemel2010a,Zemel2010b,Deshpande2006,Deshpande2007,Walcott2010,Nisenholz2014,Raab2012,Schaller2010,Kraikivski2006,Swaminathan2009,Soares2011,Alvarado2014,Deshpande2012,Deshpande2008,Rens2018,Pathak2008,Ronan2014,Reymann2010,Letort2015,Liu2008}. Yet, the agreement between our theoretical and experimental results suggests that, on the scale of the whole cell, the myriad of complex mechanisms that govern the dynamics of the stress fibers in adherent cells can be effectively described in terms of simple entropic mechanisms, as those at the heart of the physics of liquid crystals.  

In addition, our analysis demonstrates that chiral symmetry breaking can originate from the natural interplay between the orientation of the stress fibers and the shape of the cell. A more detailed investigation of this mechanism is beyond the scope of this study, but will represent a challenge in the near future with the goal of shedding light on the fascinating examples of chiral symmetry breaking observed both in single cells \cite{Tee2015} and tissues \cite{Duclos2018}. 

In the future, we plan to use our model to investigate the mechanics of cells adhering to micropatterned substrates that impose reproducible cell shapes \cite{Thery2010}, with special emphasis to the interplay between cytoskeletal anisotropy and the geometry of the adhesive patches \cite{Thery2006a}. These systems are not new to theoretical research, but previous studies have focused on either the cytoskeleton \cite{Pathak2008} or on cell shape \cite{Albert2014}, rather then on their interaction. Furthermore, our framework could be extended to study the role of cytoskeletal anisotropy in cell motility, for instance by taking into account the dynamics of focal adhesions \cite{Deshpande2008,Rens2018}, biochemical pathways in the actin cytoskeleton \cite{Maree2012}, or cellular protrusions and retractions \cite{Segerer2015}. Finally, our approach could be extended to computational frameworks such as vertex models, cellular Potts models, or phase field models \cite{Albert2016a}, and could provide a starting point for exploring the role of anisotropy in multicellular environments such as tissues~\cite{Eastwood1998,VanderSchaft2011,Wartlick2011,vanOers2014,Santos-Oliveira2015,Vijayraghavan2016,Saw2017,Barton2017}.

\section*{author contributions}

L.G. designed this study with the help of T.S., R.M.H.M. and E.H.J.D. K.S. and L.G. developed the theoretical framework. K.S. and J.E. developed the code, and J.E. performed the simulations. J.E. and W.P. carried out the analysis of the experimental data. K.S. and L.G. wrote the article. All authors commented on the manuscript.

\acknowledgements

This work was supported by funds from the Netherlands Organisation for Scientific Research (NWO/OCW), as part of the Frontiers of Nanoscience program (L.G.), the Netherlands Organization for Scientific Research (NWO-FOM) within the program on Barriers in the Brain (W.P. and T.S.; No. FOM L1714M), the Netherlands Organization for Scientific Research (NWO-ALW and NWO-ENW) within the Innovational Research Incentives Scheme  (R.M.H.M.; Vidi 2010, No. 864.10.009 and Vici 2017, No. 865.17.004), and the Leiden/Huygens fellowship (K.S.).

\appendix

\section{Angular coordinates of the adhesion sites}

With reference to the schematic representation of Fig. \ref{fig2}, the coordinates of the center of the ellipse can be expressed as:
\begin{subequations}\label{eq:center}
\begin{align}
x_{c} &= \frac{d}{2}\cos\phi-\gamma\rho\sin\phi\;,\\
y_{c} &= \frac{d}{2}\sin\phi+\rho\cos\phi\;,
\end{align}
\end{subequations}
where the length scale $\rho$ is defined in Eq. (\ref{eq_rho}). From Eqs. \eqref{eq:center}, standard algebraic manipulations allow us to express the angular coordinate $\psi$ of the adhesion points in the frame of the ellipse (Fig. \ref{fig2}a), namely
\begin{subequations}\label{eq:angles}
\begin{align}
\tan \psi_0 &= \frac{d\sin \phi + 2\rho\cos\phi}{d\cos \phi - 2\gamma\rho\sin\phi}\;,\\[10pt]
\tan \psi_1 &= \frac{d\sin \phi-2\rho\cos\phi}{d\cos \phi+ 2\gamma\rho\sin\phi}\;.
\end{align}
\end{subequations}

\section{Numerical methods}
\label{numerical_scheme}

The starting point of our numerical simulations are the positions of the adhesion sites. These are directly determined from the experiments by detecting the pillars along the cell contour that are subject to the largest traction forces (see Ref. \cite{Pomp2018} for more details). These adhesion sites are fixed during the simulation. Cellular arcs are parameterized in terms of a discrete number of vertices connected by straight edges in a chain-like manner. The bulk of the cell, representing the cytoskeleton, is defined as the region enclosed by the cell edge, and is discretized as a regularly spaced two-dimensional square lattice with $\bm{\hat{Q}}$ defined at every lattice point.

At $t=0$ the cell consists of an irregular polygon enclosing a random configuration of the nematic tensor. Evidently, this bears no resemblance to a real cell, but reduces the risk of a possible bias in the final configuration. The time-integration is performed by alternating updates of the cell contour and of the bulk nematic tensor. This process is iterated until both the cell edge and the cell bulk reach a steady-state. Our code is available upon request.

The configurations of the cytoskeleton in Figs. \ref{fig:phase_diagram}, \ref{fig_aspect_ratio}, \ref{fig_cell_example} and \ref{fig_all_cells} have been visualized with Mathematica Version 11.3 (Wolfram Research, Champaign, IL) using the line integral convolution tool.  

\subsection{Numerical implementation of the cell bulk} 
\begin{figure}[t!]
\centering
\includegraphics[width=\columnwidth]{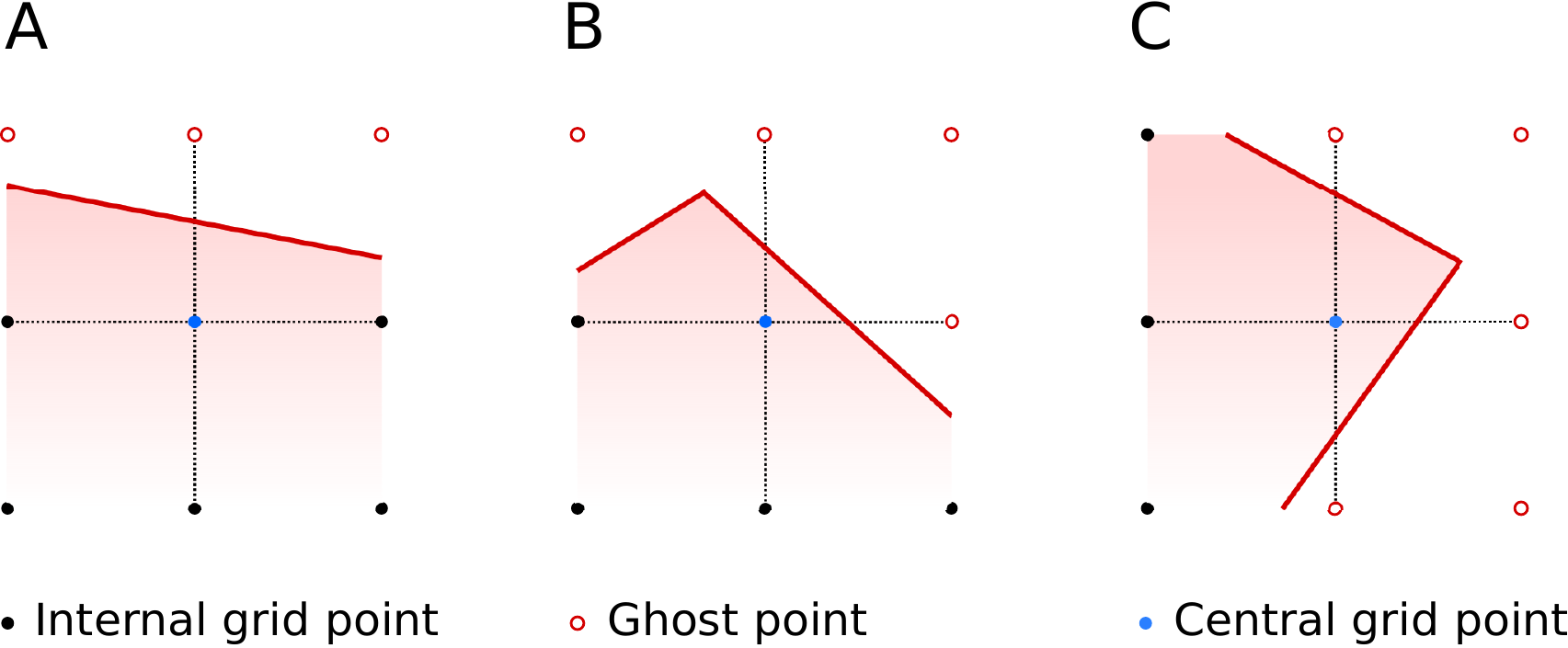}
\caption{Schematic overview of the three geometrical situations described in Appendix \ref{numerical_scheme}. (A) There is a single ghostpoint on the $x-$ or $y-$axis. (B) There are two ghost points, one on each axis. (C) There are two ghost points on the same axis and possibly a third one on the other axis.}
\label{fig_ghostpoints}
\end{figure}

Eq. \eqref{eq:eom2b} is numerically solved via a finite-difference scheme. Time integration is performed using the forward Euler method, whereas spatial derivatives are calculated using the centered difference approximation. In order to calculate derivatives at lattice points located in proximity of the edge, we use the boundary conditions, specified in Eq. (\ref{eq:bc}), to express the values of $Q_{xx}$ and $Q_{xy}$ in a number of {\em ghost points} located outside the cells. This is conveniently done upon identifying three possible scenarios, illustrated in Fig. \ref{fig_ghostpoints}. {\em 1)} There is a single ghost point on the $x-$ or $y-$axis (Fig. \ref{fig_ghostpoints}A). {\em 2)} There are two ghost points, one on each axis  (Fig. \ref{fig_ghostpoints}B). {\em 3)} There are two ghost points on the same axis and possibly a third one on the other axis (Fig. \ref{fig_ghostpoints}C). In the following, we explain how to address each of these cases.

{\em 1)} Using the centered difference approximation for the first derivative yields the following expression of the nematic tensor at a ghost point located at $(x\pm\Delta x,y)$ or $(x,y\pm\Delta y)$, with $\Delta x =\Delta y$ the lattice spacing:
\begin{subequations}
\label{eq:ghostpoint3}
\begin{multline}\label{eq:ghostpoint3a}
Q_{ij}(x\pm\Delta x,y) = Q_{ij}(x\mp\Delta x,y) \\ \pm 2\Delta x\,\partial_{x} Q_{ij}(x,y)\;,
\end{multline}
\begin{multline}\label{eq:ghostpoint3b}
Q_{ij}(x,y\pm\Delta y) = Q_{ij}(x,y\mp\Delta y) \\ \pm 2\Delta y\,\partial_{y} Q_{ij}(x,y)\;.
\end{multline}
\end{subequations}
The derivative with respect to $x$ in Eq. (\ref{eq:ghostpoint3a}) can be calculated from Eq. \eqref{eq:bc}, upon taking $\bm{N}=\pm\bm{\hat{x}}$, where the plus (minus) sign correspond to a ghost point located on the left (right) of the central edge point. Thus $\bm{N}\cdot\nabla Q_{ij}=\pm\partial_{x}Q_{ij}$. Analogously, the derivative with respect to $y$ in Eq. (\ref{eq:ghostpoint3b}), is approximated as $\bm{N}\cdot\nabla Q_{ij}=\pm\partial_{y}Q_{ij}$, where the plus (minus) sign corresponds to a ghost point located below (above) the central edge point. Combining this with Eq. \eqref{eq:bc}, yields:
\begin{subequations}
\label{eq:ghostpoint5}
\begin{multline}
Q_{ij}(x\pm\Delta x,y) = Q_{ij}(x\mp\Delta x,y)\\
-4\Delta x\,\frac{W}{K} \left[Q_{ij}(x,y) - Q_{0,ij}(x,y)\right]\;,
\end{multline}
\begin{multline}
Q_{ij}(x,y\pm\Delta y) = Q_{ij}(x,y\mp\Delta y)\\
-4\Delta y\,\frac{W}{K} \left[Q_{ij}(x,y) - Q_{0,ij}(x,y)\right]\;.
\end{multline}
\end{subequations}
The tensor $Q_{0,ij}$ is evaluated via Eq. \eqref{eq_q0} using the local orientation of the cell edge.

{\em 2)} If a given lattice point is linked to ghost points in both the $x-$ and $y-$directions, we evaluate equation \eqref{eq:ghostpoint5} for both directions independently as explained in the previous paragraph.

{\em 3)} If a given lattice point is linked to two ghost points in either the $x-$ or $y-$direction, we employ a forward or backward finite difference approximation for the first spatial derivative of $Q_{ij}$ to evaluate $Q_{ij}$ at the ghost points. This yields:
\begin{subequations}
\label{eq:ghostpoint8}
\begin{multline}
Q_{ij}(x\pm\Delta x,y) = Q_{ij}(x,y) \\ -2\Delta x\frac{W}{K} \left[Q_{ij}(x,y) - Q_{0,ij}(x,y)\right]\;,
\end{multline}
\begin{multline}
Q_{ij}(x,y\pm\Delta y) = Q_{ij}(x,y) \\ -2\Delta y\frac{W}{K} \left[Q_{ij}(x,y) - Q_{0,ij}(x,y)\right]\;.
\end{multline}
\end{subequations}
Finally, if the given lattice point is also linked to a ghost point on the other axis, this is evaluated independently using Eq. \eqref{eq:ghostpoint5}. 

The lattice points enclosed by the cells continuously change during the course of a simulation, as a consequence of the relaxation of the cell shape. In case of inclusion of a new lattice point that was previously located outside the cell, the associated $Q_{xx}$ and $Q_{xy}$ values are generated by averaging over the nearest neighbours (horizontally and vertically, not diagonally) that were inside the cell during the previous time step. 

\subsection{Numerical implementation of the cell edge} 

To calculate the line tension $\lambda$, Eq. \eqref{eq:lambda} is discretized as follows:
\begin{equation}
\label{eq_lambda_discrete}
\lambda_k = \lambda_0 - \alpha_0 \sum_{n = 1}^{k} \Delta s_n\,\bm{T}_n \cdot \left\langle\bm{\hat{Q}}_n\right\rangle \cdot \bm{N}_n\;, \quad k=1,2\ldots\,N_{\rm arc}\;,
\end{equation}
where $\lambda_0$ is the line tension at the adhesion site at $s=0$ (position $\bm{r}_0$) and $\lambda_k$ is the line tension at vertex $k$ (position $\bm{r}_k$). $N_{\rm arc}$ is the total number of vertices in which cellular arcs are discretized, and $\lambda_{N_{\rm arc}}$ represents the line tension at the other adhesion site. Furthermore, $\Delta s_{n} = |\bm{r}_{n}-\bm{r}_{n-1}|$, $\bm{T}_{n} = (\bm{r}_{n}-\bm{r}_{n-1})/\Delta s_{n}$, $\bm{N}_n =\bm{T}_n^{\perp} $ and 
\begin{equation}
\left\langle\bm{\hat{Q}}_n\right\rangle =\frac{\bm{\hat{Q}}_n + \bm{\hat{Q}}_{n-1}}{2}\;,
\end{equation}
with $\bm{\hat{Q}}_n$ and $\bm{\hat{Q}}_{n-1}$ the nematic tensor at the vertices $n$ and $n-1$. These are set equal to $\bm{\hat{Q}}$ at the closest bulk lattice point inside the cell among the four points delimiting the unit cell containing the edge vertices $n$ and $n-1$ respectively. If none of these is inside the cell, we set $Q_{xx,n}=Q_{xy,n}=0$. The quantity $\lambda_0$ is calculated in such a way that the minimal $\lambda$ value along an arc equates the input parameter $\lambda_{\min}$, representing the minimal tension withstood by the cortical actin. 

Next, the position of the vertices $\bm{r}_{k}$, $k=0, 1\ldots\, N_{\rm arc}$ is updated upon integrating Eq. \eqref{eq:eom2a} using the forward Euler method. The curvature and normal vector at vertex $k$, $\kappa_{k}$ and $\bm{N}_{k}$, are found by constructing a circle with radius $R$ through vertices $k -1$, $k$, and $k +1$. The vector from vertex $k$ to the center of the circle is then equated to $\pm R \bm{N}_k$, with the sign such that $\bm{N}_k$ is an inward pointing normal vector, and $\kappa_k = \pm 1/R$, with a negative sign for a concave shape and a positive sign for a convex shape. Along each arc, $\bm{r}_{0}$ and $\bm{r}_{N_{\rm arc}}$ represent the positions of the adhesion sites and are kept fixed during simulations.

\section{Experimental data analysis}
\label{experimental_methods}

The coarse-grained experimental data displayed in Figs. \ref{fig_cell_example}B and \ref{fig_all_cells}F-J are obtained using the following four-step procedure.

{\em 1)} Epithelioid GE11 and fibroblastoid GD25 cells expressing either $\upalpha5\upbeta1$ or $\upalpha \rm v\upbeta3$ (GD$\upbeta$1, GD$\upbeta$3, GE$\upbeta$1 and GE$\upbeta$3) \cite{Balcioglu2015} are cultured on microfabricated elastomeric pillar arrays \cite{Tan2003,Trichet2012,VanHoorn2014}. The resulting cells are imaged using spinning disk confocal microscopy and displayed in Figs. \ref{fig_cell_example}A and \ref{fig_all_cells}A-E. For details, see Ref. \cite{Pomp2018}. 

{\em 2)} The locations of the cell interior and the cell edge are found by applying a low-pass filter on the images using Matlab. The interior of the cell is then sampled by overlaying a square lattice of $512 \times 512$ pixels ($1\,{\rm pixel} = 0.138 \times 0.138$ $\mu{\rm m}^{2}$) on the microscope field-of-view (Figs. \ref{fig_cell_example}A and \ref{fig_all_cells}A-E).

{\em 3)} For all pixels that are inside the cell, we calculate the nematic tensor using ImageJ with the OrientationJ plugin \cite{OrientationJ} in the following way. Given the intensity $I(x_0,y_0)$ of the image at the point $(x_0,y_0)$, we define the symmetric $2\times 2$ matrix $\bm{\hat{J}} =\langle\nabla I \nabla I\rangle$, where $\langle\cdots \rangle = \int w(x,y){\rm d}x\,{\rm d}y\,(\cdots)$ represents a weighted average with $w (x,y) $ a Gaussian with a standard deviation of five pixels ($0.69\:\mu$m) centered at $(x_0,y_0)$. The $\bm{\hat{J}}$ matrix can be expressed as:
\begin{equation}
\label{eq_decomposition}
\bm{\hat{J}} 
= \left(\Lambda_{\min} -\Lambda_{\max}\right) \left(\bm{e}_{\min}\bm{e}_{\min}-\frac{1}{2}\bm{\hat{I}}\right)
+ \frac{\Lambda_{\max}+\Lambda_{\min}}{2}\,\bm{\hat{I}}\;,
\end{equation}
where $\Lambda_{\mathrm{max}}$ and $\Lambda_{\mathrm{min}}$ are the largest and smallest eigenvalues of $\bm{\hat{J}}$, $\bm{e}_{\mathrm{min}}$ the eigenvector corresponding to $\Lambda_{\mathrm{min}}$, and $\bm{\hat{I}}$ the two-dimensional identity matrix. The $\bm{\hat{J}}$ matrix is then used to estimate the average stress fiber direction $\bm{u}$:
\begin{equation}
\label{eq_assumption}
\frac{\langle\nabla I \nabla I\rangle}{\langle |\nabla I|^2\rangle} =  \bm{\hat{I}} -\langle\bm{u}\bm{u}\rangle\;.
\end{equation}
Here, the quantity $\bm{\hat{I}} -\langle\bm{u}\bm{u}\rangle$ reflects that the largest gradients in intensity are perpendicular to the orientation of the stress fibers and $\langle |\nabla I|^2\rangle =\tr \bm{\hat{J}} =\Lambda_{\max} +\Lambda_{\min}$. Combining Eqs. (\ref{eq_decomposition}) and \eqref{eq_assumption}, we obtain
\begin{equation}
\label{eq_order_parameter}
\left\langle \bm{u}\bm{u} -\frac{1}{2}\,\bm{\hat{I}} \right\rangle
= \frac{\Lambda_{\max} -\Lambda_{\min}}{\Lambda_{\max} +\Lambda_{\min}}\,\left(\bm{e}_{\min}\bm{e}_{\min} -\frac{1}{2}\,\bm{\hat{I}}\right)\;.
\end{equation}
Comparing this with the definition of the nematic tensor \cite{Gennes_book}:  
\begin{equation}
\bm{\hat{Q}} 
= \left \langle \bm{u}\bm{u} -\frac{1}{2}\,\bm{\hat{I}}\right\rangle 
= S \left(\bm{n}\bm{n} -\frac{1}{2}\,\bm{\hat{I}}\right)\;,
\end{equation}
we find:
\begin{equation}
S = \frac{\Lambda_{\max}-\Lambda_{\min}}{\Lambda_{\max}+\Lambda_{\min}}\;,\qquad
\bm{n} = (\cos \theta_{\mathrm{SF}},\sin \theta_{\mathrm{SF}}) =\bm{e}_{\mathrm{min}}\;.
\end{equation}
{\em 4)} The data are further coarse-grained in blocks of $8\times 8$ pixels corresponding to regions of size $1.104 \times 1.104\,\mu m^{2}$ in real space. This results in a new $64\times 64$ lattice. The value of the nematic tensor in the new coarse-grained pixels is obtained from an average over those of the original $8\times 8$ pixels located inside the cell. In turn, the coarse-grained pixels are considered inside the cell if more than half of the original pixels were inside the cell.

\bibliographystyle{ieeetr}

\end{document}